\documentclass[amsmath,amssymb,aps,prl,reprint,superscriptaddress]{revtex4-2}

\usepackage[pdftex]{graphicx}
\usepackage[dvipsnames,svgnames]{xcolor}

\usepackage[
    colorlinks=true,
    linkcolor=DarkRed,
    citecolor=ForestGreen,
    urlcolor=MediumBlue,
    pdfstartview=FitH,
    bookmarks=false,
    pdfpagemode=UseNone
]{hyperref}

\usepackage{placeins}
\usepackage{dcolumn}
\usepackage{comment}
\usepackage[title]{appendix}
\usepackage{stix}

\begin{document}
\title{Ultrafast dynamics of excitons in black phosphorus}

\author{Geoffroy Kremer}
\affiliation{Universit\'e de Lorraine, CNRS, IJL, F-54000 Nancy, France}
\affiliation{Department of Physics, University of Fribourg, Chemin du Mus\'ee 3, 1700 Fribourg, Switzerland}

\author{Juan F. P. Mosquera}
\affiliation{PSI Center for Scientific Computing, Theory and Data, Paul Scherrer Institute, 5232 Villigen PSI, Switzerland}
\affiliation{Department of Physics, University of Fribourg, Chemin du Mus\'ee 3, 1700 Fribourg, Switzerland}

\author{Jo\"el Morf}
\affiliation{Department of Physics, University of Fribourg, Chemin du Mus\'ee 3, 1700 Fribourg, Switzerland}

\author{Aymen Mahmoudi}
\affiliation{Department of Physics, University of Fribourg, Chemin du Mus\'ee 3, 1700 Fribourg, Switzerland}

\author{Fr\'ed\'eric Chassot}
\affiliation{Department of Physics, University of Fribourg, Chemin du Mus\'ee 3, 1700 Fribourg, Switzerland}

\author{Viktor Christiansson}
\affiliation{Institute of Solid State Physics, TU Wien, Vienna, Austria}

\author{Maxime Rumo}
\affiliation{Department of Physics, University of Fribourg, Chemin du Mus\'ee 3, 1700 Fribourg, Switzerland}

\author{Manuele Balestra}
\affiliation{Department of Quantum Matter Physics, University of Geneva, CH-1211 Geneva, Switzerland}

\author{Fabian O. von Rohr}
\affiliation{Department of Quantum Matter Physics, University of Geneva, CH-1211 Geneva, Switzerland}

\author{Philipp Werner}
\affiliation{Department of Physics, University of Fribourg, Chemin du Mus\'ee 3, 1700 Fribourg, Switzerland}

\author{Michael Sch\"uler}
\affiliation{PSI Center for Scientific Computing, Theory and Data, Paul Scherrer Institute, 5232 Villigen PSI, Switzerland}
\affiliation{Department of Physics, University of Fribourg, Chemin du Mus\'ee 3, 1700 Fribourg, Switzerland}

\author{Claude Monney}
\affiliation{Department of Physics, University of Fribourg, Chemin du Mus\'ee 3, 1700 Fribourg, Switzerland}

\renewcommand{\vec}{\mathbf}
\newcommand{\gvec}{\boldsymbol}

\begin{abstract}
Excitons are key quasi-particles determining the optical properties of solids. As such, they can be utilized to coherently control the electronic structure of materials using optical femtosecond pulses. Identifying the decoherence mechanism during the early non-equilibrium dynamics is crucial to achieve light‑induced band‑structure engineering in semiconductors. Here, we generate excitons in the direct band gap semiconductor black phosphorus with a resonant mid-infrared photoexcitation. Using time- and angle-resolved photoemission spectroscopy, we track their complex ultrafast dynamics on the few-picosecond time scale. We develop a quantum-kinetic theoretical framework to model the decoherence of excitons into dark excitons via phonon scattering. By combining simulation and experiment, we quantify key parameters describing the early dynamics of the excitons. Our work highlights phonon-mediated intravalley scattering as a fundamental limitation for coherent exciton phenomena in single-valley semiconductors. \\

\end{abstract}

\flushbottom
\maketitle

\thispagestyle{empty}

\section*{Introduction}

Excitons -- electron-hole pairs bound by the Coulomb interaction -- govern the optical response of semiconductors and play a central role in a broad class of nonequilibrium phenomena, from light harvesting to ultrafast control~\cite{wang_colloquium_2018}.  
Their coherent superposition states, in which the exciton wavefunction carries a well-defined quantum mechanical phase, are of particular current interest across several fields. In quantum information science, exciton coherence underpins proposals for optically addressable qubits and ultrafast all-optical logic~\cite{gucci_encoding_2026}. In condensed matter physics, proposals for Floquet engineering of exciton bands, the formation of hybrid light-matter polariton states, and the optically driven control of excitonic order all rely critically on the ability to generate and sustain coherent exciton populations over timescales long enough to manipulate them~\cite{kasprzak_boseeinstein_2006,Kobayashi2023,Pareek2026,Chan2023}. In all of these examples, the key ingredient is the exciton coherence, and its survival is the central limiting factor.

Answering this question quantitatively has proved difficult. On the theoretical side, substantial progress has been made: excitonic absorption spectra including the linewidth due to exciton-phonon coupling can be computed~\cite{antonius_theory_2022,chan_exciton_2023,lechifflart_first-principles_2023}, and quantum-kinetic frameworks extending the semiconductor Bloch equations to include exciton-phonon scattering are able to describe real-time exciton dynamics~\cite{kira_many-body_2006,chen_first-principles_2022,Stefanucci_ExcitonicBlochEquations_2025,chan_excitonphonon_2024,chan_exciton_2025,mittenzwey_excitonic_2025,dogadov_dissecting_2026}. On the experimental side, however, accessing coherence dynamics directly remains a challenge: optical spectroscopies are sensitive to the macroscopic polarization and convolve population and coherence dynamics in ways that are difficult to disentangle. A direct, momentum-resolved window onto the full exciton dynamics — bright and dark states alike — is needed.

Time- and angle-resolved photoemission spectroscopy (trARPES) has emerged as a uniquely powerful tool in this context~\cite{boschini_time-resolved_2024}. By directly tracking the momentum-resolved distribution of photoexcited carriers as a function of pump-probe delay, trARPES grants access to exciton dynamics in ways that are complementary to and often beyond what optical spectroscopies can reveal~\cite{man_experimental_2021, Dong2021, reutzel_probing_2024}.
Most works to date has focused on transition-metal dichalcogenides (TMDCs), where the multivalley electronic structure introduces a rich interplay between spin, valley, and exciton degrees of freedom \cite{Bertoni2016, Dong2021,Chen2022,Helmrich2021,Schmitt2022,Christiansen2019,Wallauer2021}. In these materials, intervalley scattering — mediated by phonons or disorder — plays a dominant role in exciton relaxation and dephasing, efficiently transferring population from optically bright states at zero center-of-mass momentum to momentum-forbidden dark exciton states on femtosecond to picosecond timescales. While this physics is now relatively well characterized, the multivalley complexity also obscures more elementary mechanisms of exciton decoherence that are intrinsic to simpler, single-valley semiconductors.

Black phosphorus (BP) provides a compelling alternative platform to isolate and study these elementary processes. With a single relevant valley at the $\bar{\Gamma}$ point of a rectangular surface Brillouin zone, BP offers a unique opportunity to isolate exciton dynamics without the complications of intervalley scattering. Bulk BP hosts excitons with pronounced in-plane anisotropy, strong light-matter coupling, and a direct band gap that depends on layer thickness across the infrared to visible range~\cite{tran_layer-controlled_2014, zhang_determination_2018, wang_highly_2015}. Despite these favorable properties and growing experimental interest in its nonequilibrium response~\cite{zahn_anisotropic_2020, montanaro_anomalous_2022,shen_ultrafast_2025,dendzik_ultrafast_2026}, the microscopic dynamics of coherent excitons in BP — their formation, dephasing, and conversion into dark states — have remained largely unexplored.

\begin{figure*}[ht]
\centering
\includegraphics[width=\linewidth]{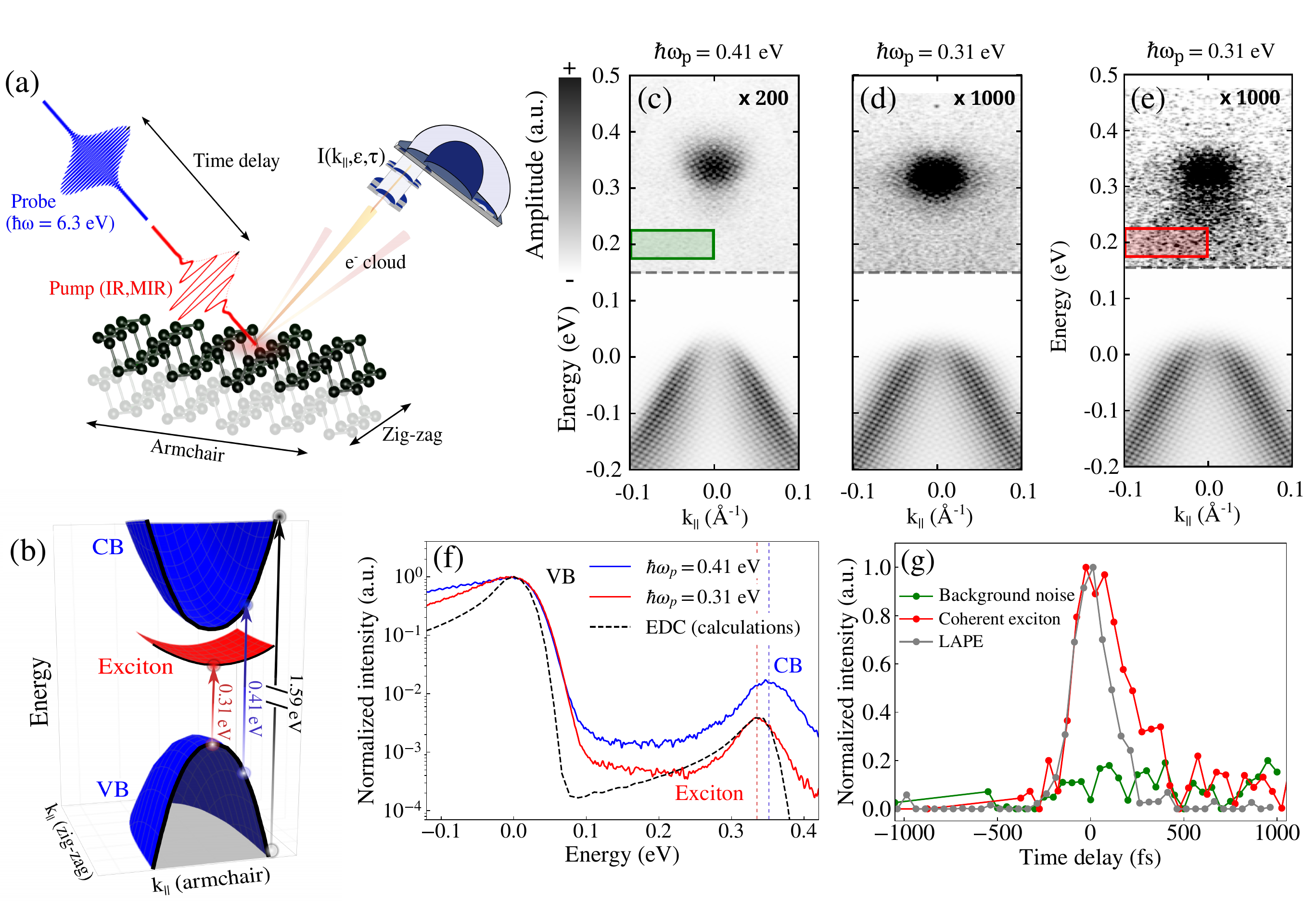}
\caption{\textbf{Observation of excitons in bulk black phosphorus.} 
(a) Sketch of the experimental setup. Electrons are excited by either a IR or MIR pulse and photoemitted by a $\hbar\omega_\mathrm{pr} = 6.3$~eV pulse. 
(b) Dispersion of the valence band (VB), conduction band (CB), and the exciton state in the band gap.
The arrows show the different excitation energies used in the experiments: $\hbar \omega_p = 1.55$ eV (black), $\hbar \omega_p = 0.41$ eV (blue) and $\hbar \omega_p = 0.31$ eV (red). The latter photon energy is resonant with the coherent exciton at $\mathbf{Q}=0$. (c) Pump-probe ARPES spectrum acquired with pump photon energy $\hbar\omega_\mathrm{p} = 0.41$~eV along the armchair direction, integrated from 500 fs to 10 ps. The data are symmetrized around $k_\parallel=0$ \AA$^{-1}$.
(d) Pump-probe ARPES spectrum as in (c), but with pump photon energy $\hbar\omega_\mathrm{p} = 0.31$~eV. A downwards energy shift of the dominant feature in the unoccupied states is observed compared to (c).  
(e) Pump-probe ARPES spectrum as in (d), but integrated from -200 fs to +200 fs, showing a transient signal inside the band gap. 
(f) Energy distribution curves (EDC) extracted from panels (c) and (d) in the momentum range [-0.05, 0.05] \AA$^{-1}$, together with the calculated EDC (see main text). The black, red and blue dashed lines indicate the energy position of the VB maximum, exciton peak and CB, resp..
(g) Transient photoemission intensity extracted from the region of interest (ROI) indicated in panel (e) for the $\hbar\omega_\mathrm{p}=0.31$~eV pump, corresponding to the coherent exciton (with $s$-polarized pump light) and the laser-assisted photoemission (LAPE) (with $p$-polarized pump light). Both curves are normalized to their maximum intensity. The green curve, the intensity of which is scaled for visibility, represents the background signal from the ROI indicated in panel (c) for the $\hbar\omega_\mathrm{p}=0.41$~eV pump. 
}
\label{fig1}
\end{figure*}

Here we address this gap by combining trARPES with a quantum-kinetic theoretical framework designed for quantitative comparison with the experiment. Using a tunable mid-infrared (MIR) pump laser we can selectively excite excitons, directly observe the formation of coherent excitons at zero center-of-mass momentum, and track their redistribution in energy-momentum space on a few-picosecond timescale. A key feature of our work is the modeling philosophy: rather than computing the exciton-phonon coupling from first principles, we develop a comprehensive simulation of the trARPES response that incorporates the full complexity of photoemission -- photoemission matrix elements, out-of-plane momentum ($k_z$) broadening, laser pulse parameters, the \emph{ab initio} exciton dispersion, and the real-time quantum kinetics of exciton populations and coherences. This level of quantitative modeling, which we believe has not previously been achieved for exciton dynamics measured by trARPES, enables us to use the experimental data not just as a benchmark but as a precise constraint on the underlying microscopic parameters. By optimizing the agreement between simulation and experiment across multiple energy-momentum regions simultaneously, we extract the exciton-phonon coupling strength and determine the coherence lifetime with quantitative reliability.

Our results reveal that intravalley exciton-phonon scattering suppresses the coherence of the bright exciton within a few tens of femtoseconds, converting it into a long-lived population of dark, finite-momentum excitons that persists for tens of picoseconds. This scattering is driven by phonon absorption rather than emission. Comparison with measurements at higher pump photon energies reveals markedly different relaxation pathways and timescales, underscoring the sensitivity of the nonequilibrium dynamics to the initial excitation. Our findings establishes BP as a model system for disentangling coherent and incoherent exciton dynamics and highlights intravalley scattering as a fundamental limitation for coherent exciton phenomena in single-valley semiconductors.  

\section*{Results}
\textbf{Observation of resonantly populated excitons}

We perform trARPES in the pump-probe scheme to photoexcite excitons in BP and follow their ultrafast dynamics. In our experiment, we use variable pump photon energies and a probe photon energy of 6.3 eV [see Fig.~\ref{fig1}(a)]. Note that BP is highly anisotropic with a rectangular surface Brillouin zone \cite{golias2016disentangling}. We focused on the most dispersive direction and show in Fig.~\ref{fig1} trARPES data obtained along the armchair direction near the $\bar{\Gamma}$ point. In graph Fig.~\ref{fig1}(c), the data are averaged for time delays between 500 fs and 10 ps for a photoexcitation energy of 0.41 eV, which is larger than the band gap of BP [see Fig. \ref{fig1}(b)]. At this photon energy and in this time delay range, the very bottom of the lowest conduction band (CB) is directly populated by electrons excited from the top of the valence band (VB). Both bands, demarcating a band gap of about 0.3 eV, are observed in Fig.~\ref{fig1}(c), in which the energy is referenced with respect to the maximum of the VB. We then reduce the excitation energy to 0.31 eV, comparable with the band gap size. In the corresponding trARPES data of Fig.~\ref{fig1}(d), the intensity near the position of the CB adopts a slightly different distribution in energy and momentum space. To make it more obvious, we plot in Fig.~\ref{fig1}(f) the energy distribution curves (EDC) integrated over a momentum range near the band extrema ([-0.05, 0.05] \AA$^{-1}$) for the two different photoexcitation photon energies. While the position of the leading edge of the VB matches well in both cases, we see that the peak above the VB is shifted to lower energies by 14 meV for the 0.31 eV photoexcitation (red curve), compared to the position of the CB (grey curve) for the 0.41 eV photoexcitation. This energy shift is similar to the expected exciton binding energy $E_{ex}$ in bulk BP \cite{Carre2021}. This is the first evidence that we are able to resonantly photoexcite excitons in bulk BP.

We then turn to the transient evolution of the photoemission intensity. For this purpose, ARPES data integrated from -200 fs to +200 fs are displayed in Fig.~\ref{fig1}(e). Interestingly, some signal appears in the band gap of BP, showing a downward dispersion \textit{below} the CB. This is reminiscent of the dispersion expected for excitons in semiconductors at early times \cite{rustagi_photoemission_2018,Dong2021}. We then plot in Fig.~\ref{fig1}(g) the photoemission signal \textit{inside} the band gap, at about 0.2 eV above the VB maximum (VBM). While the transient signal acquired with above-band-gap photoexcitation (0.41 eV, green curve) is mainly noise with slow dynamics coming from the tail of the VB, we see a drastically different behavior in the case of near-band-gap photoexcitation (0.31 eV, red curve). An asymmetric peak appears near $t_0$. This in-gap signal, featuring a short, but non-zero lifetime, is our second evidence of a transient excitonic signal. Up to now, $s$-polarized photons were used for photoexcitation. Note that we have carefully optimized our probe-laser polarization waveplate angle by minimizing the laser-assisted photoemission (LAPE) signal on the reference material Bi$_2$Se$_3$ (see SM).
We overlay on the same graph the cross-correlation signal (gray curve) obtained by rotating the photoexcitation polarization to $p$, for which the LAPE signal is dominating. This highlights the excitonic contribution that persists at times later than 500 fs.

\begin{figure*}[ht]
\centering
\includegraphics[width=\linewidth]{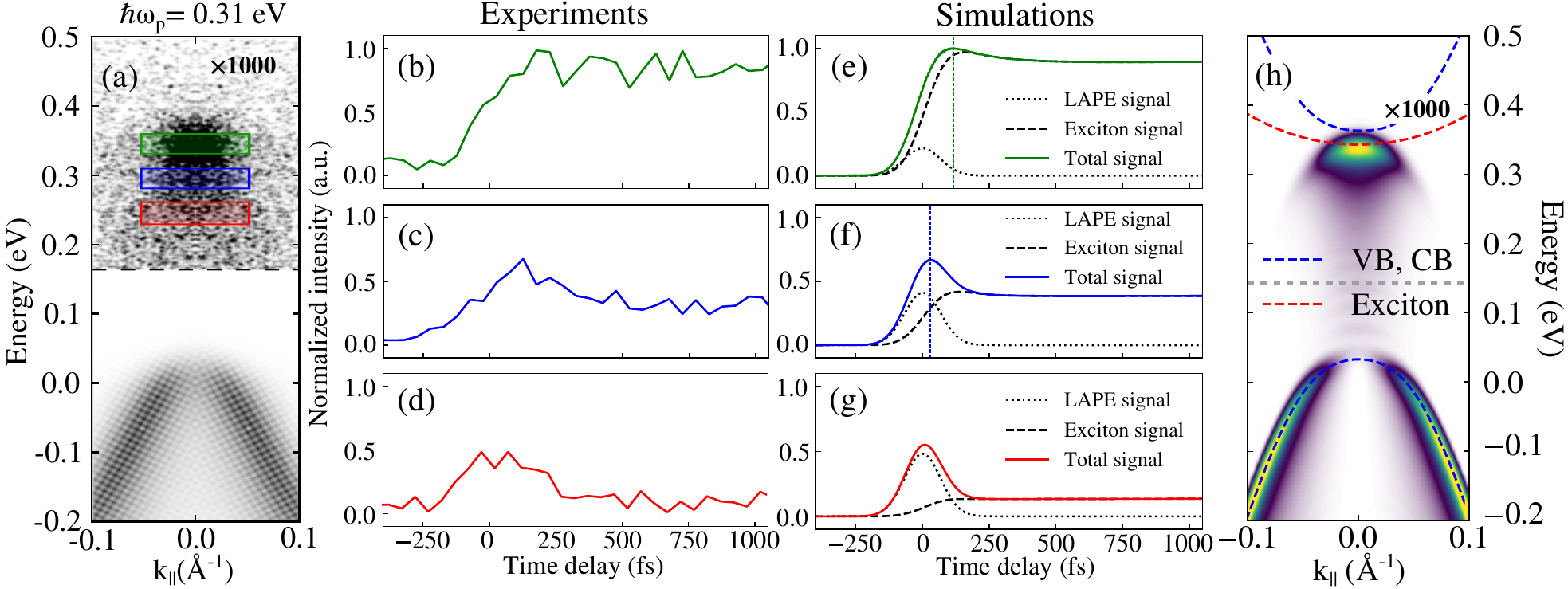}
\caption{\textbf{Transient dynamics of excitons.} (a) ARPES spectrum acquired with near-band-gap photoexcitation (0.31 eV) for overlapping pump and probe pulses. The colored boxes indicate the regions of interest (ROI) at which the transient dynamics are extracted in panels (b)-(g). (b)-(d) Transient photoemission intensity in different ROI, as indicated in panel (a). The curves are normalized to the maximum value of the ROI with the highest intensity. (e)-(g) Corresponding simulated transient photoemission intensity in the same ROI indicated in panel (a), including both exciton and LAPE contributions. (h) Simulated trARPES intensity for overlapping pump and probe pulses. The dashed lines mark the dispersion of the VB and CB, as well as the exciton dispersion. The ROI for the simulated spectrum are equivalent to the ones in panel (a). The spectra in panel (a) has been scaled by a factor of 1000 above the gray dashed line as well as the exciton's signal in panel (h). }
\label{fig:trARPES_exp_theory}
\end{figure*}

Having established our capability to resonantly photoexcite excitons in bulk BP, we now inspect their ultrafast dynamics in more detail. In Fig.~\ref{fig:trARPES_exp_theory} we present the transient ARPES intensity extracted from different regions of interest (ROI) at different energies above the VB [see Fig.~\ref{fig:trARPES_exp_theory}(a) for their positions]. The three curves displayed in graphs \ref{fig:trARPES_exp_theory}(b) to \ref{fig:trARPES_exp_theory}(d) show a distinct evolution as their respective energy increases towards the CB minimum \textit{from below}. The asymmetric peak at $t_0$ vanishes, giving way to a rounded step function with a delayed maximum. We attribute the peak at $t_0$ to the resonant formation of bright excitons with center-of-mass momentum $\vec{Q}=0$. These bright excitons scatter with phonons into dark $\vec{Q}\neq 0$ excitons that appear at energies closer to the CB minimum (CBM) in our experimental data after a short time delay.
\\

\textbf{Quantum-kinetic theoretical framework}

\begin{figure*}[t]
\centering
\includegraphics[width=\linewidth]{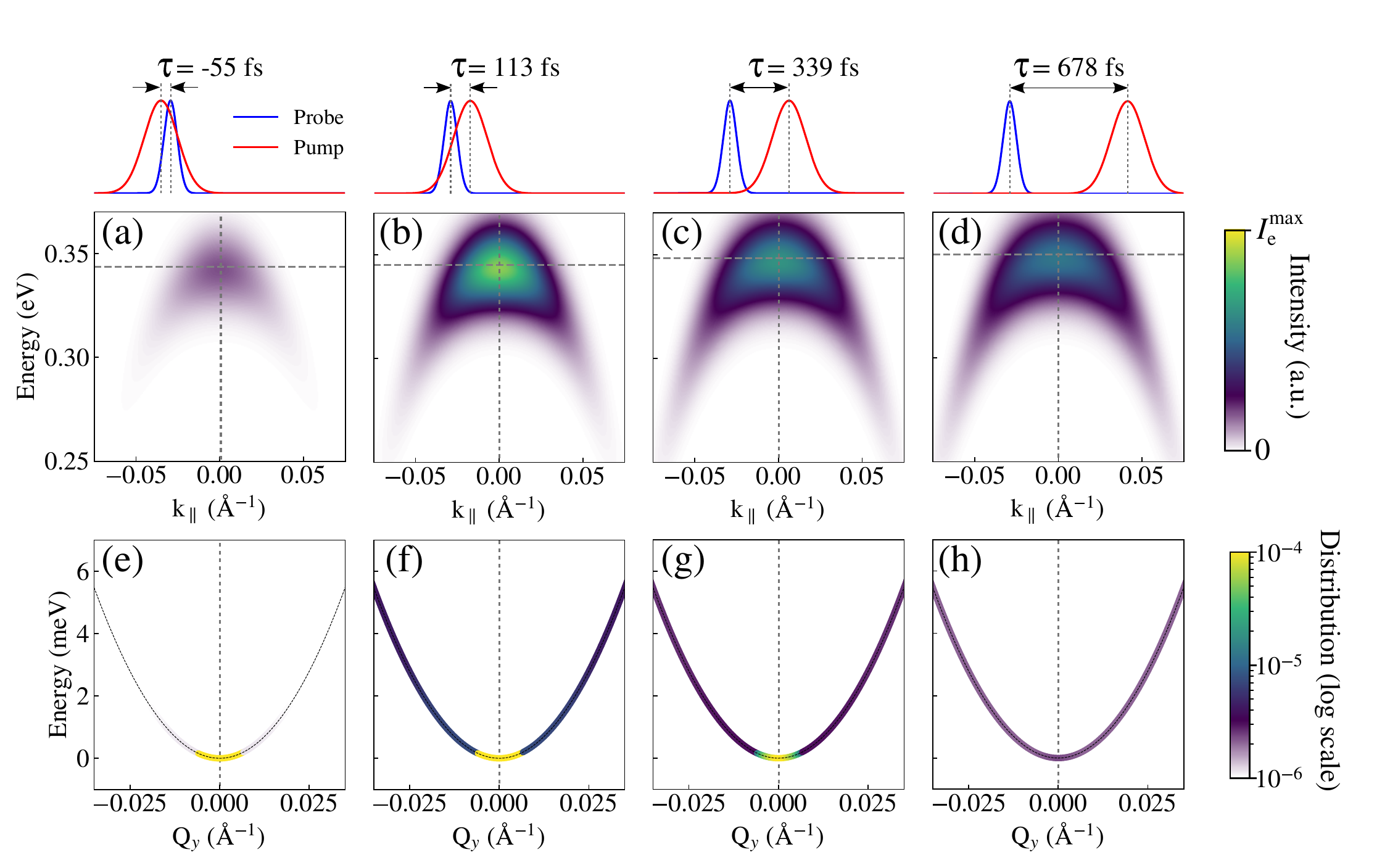}
\caption{\textbf{Exciton scattering dynamics.} 
(a)-(d) Simulated trARPES intensity for different time delays (as sketched above). 
The gray dashed lines indicate the center of mass of momentum and energy at each time delay.
(e)-(h) Exciton distribution $N_{\vec{Q}}(t)$ for time delays $\tau$ as in (a)-(d). The black dashed lines show the exciton dispersion. } 
\label{fig:exciton_population}
\end{figure*}

To gain microscopic insight into the exciton dynamics observed in our experiments, we performed simulations of the exciton dynamics in BP based on a quantum-kinetic treatment. As the first ingredient, we calculated the exciton states in bulk BP by solving the Bethe-Salpeter equation (BSE) on top of a GW quasiparticle band structure~\cite{onida_electronic_2002}. Here, we employed the method from ref.~\cite{beaulieu_berry_2024}, combining a Wannier tight-binding model~\cite{pizzi_wannier90_2020} for the quasiparticle bands with effective Coulomb interactions determined from the constrained random phase approximation (cRPA)~\cite{aryasetiawan_frequency-dependent_2004}. This approach, which is similar to those presented in refs.~\cite{bieniek_theory_2022, dias_wantibexos_2023}, allows us to use the very dense momentum grids necessary to accurately capture the small exciton binding energies. The resulting exciton band structure is shown in Fig.~\ref{fig:exciton_population}(a), revealing a binding energy of about 20 meV for the lowest-energy exciton at $\vec{Q}=0$. This value is consistent with the upper bound of 30~meV established in ref.~\cite{tran_layer-controlled_2014}. The exciton dispersion exhibits a pronounced anisotropy between the armchair and zigzag directions, reflecting the underlying electronic band structure of BP.

Solving the BSE also yields the exciton envelope functions $\Lambda^{\lambda,\vec{Q}}(\mathbf{k})$~\cite{rohlfing_electron-hole_2000}, which define the many-body wave function of an exciton in a state $\lambda$ with center-of-mass momentum $\vec{Q}$ by
\begin{align}
	\label{eq:exciton_wf}
	|\Psi^{\lambda,\vec{Q}}\rangle = \sum_{\vec{k}} \Lambda^{\lambda,\vec{Q}}(\vec{k}) \hat{c}^\dagger_{\vec{k}+\vec{Q}} \hat{v}_{\vec{k}} |\Psi_0\rangle \ .
\end{align}
Here, $\hat{c}^\dagger_{\vec{k}}$ ($\hat{v}_{\vec{k}}$) creates (annihilates) an electron in the conduction (valence) band with momentum $\vec{k}$, and $|\Psi_0\rangle$ is the electronic ground state. 

With the exciton states at hand, we turn to the ultrafast exciton dynamics. Describing the time evolution of excitons in the presence of a laser excitation and exciton-phonon interactions is a challenging many-body problem~\cite{perfetto_real-time_2023}. Here we employ a quantum-kinetic approach inspired by the semiconductor Bloch equations~\cite{lindberg_effective_1988, kira_many-body_2006} combined with the time-dependent Boltzmann equation~\cite{stefanucci_semiconductor_2024,mocatti_nonequilibrium_2025}. In this framework, we describe the exciton dynamics in terms of the exciton density matrix $\rho_{\alpha \beta}(t)$, where $\alpha, \beta \in \{0, \vec{Q}\}$ label the  ground state ($0$) and exciton states with momentum $\vec{Q}$. The diagonal elements with $\alpha=\vec{Q}$ correspond to the population $N_{\vec{Q}}(t)$ of excitons with momentum $\vec{Q}$, while the off-diagonal elements describe coherences between different exciton momentum states. In particular, the coherences $\rho_{0 \vec{Q}}(t)$ directly describe the coherent exciton polarization induced by the laser field. The time evolution of the density matrix is governed by the equation of motion
\begin{align}
	\label{eq:eom_density_matrix}
	\frac{d}{dt} \gvec{\rho}(t) =  -i [\vec{H}^\mathrm{exc}(t), \gvec{\rho}(t)] + \vec{S}(t) \ .
\end{align}
The excitonic Hamiltonian $H^\mathrm{exc}_{\alpha \beta}(t) = E_\alpha \delta_{\alpha\beta} - \vec{E}(t) \cdot \vec{D}_{\alpha\beta} $ contains the ground-state energy $E_\alpha=E_0$ (exciton energies $E_\alpha = E_x(\vec{Q})$) and the coupling to the time-dependent electric field $\vec{E}(t)$ via the excitonic dipole matrix elements $\vec{D}_{\alpha\beta}$. The electric field is modeled as a Gaussian laser pulse with the same pulse duration as the time cross-correlation in the experiments; the peak electric field $E^{(0)}$ is kept as a free parameter to match the experimental excitation density. 

The diagonal part $S_{\alpha\alpha}(t)$ of the scattering term describes exciton-phonon scattering 
processes, which we include at the level of the quantum master equation
\begin{align}
	\label{eq:scattering_term}
	S_{\vec{Q}\vec{Q}}(t) = & \frac{2\pi \gamma^2}{N_Q}\sum_{\vec{Q}^\prime} \left[ \Gamma_{\vec{Q}^\prime\rightarrow\vec{Q}} N_{\vec{Q^\prime}}(t)
	 - \Gamma_{\vec{Q}\rightarrow\vec{Q}^\prime} N_{\vec{Q}}(t) \right] \ .
\end{align} 
The scattering rates $\Gamma_{\vec{Q}\rightarrow\vec{Q}^\prime}$ are defined in terms of the acoustic phonon modes of BP and the exciton-phonon coupling matrix elements. We characterize them by the mode-averaged coupling strength $\gamma$, which is the only free parameter in our model besides $E^{(0)}$. The off-diagonal part is treated within the relaxation-time approximation $S_{\alpha\beta} = [S_{\alpha\alpha} + S_{\beta\beta}]/2$, which has been shown to accurately capture dephasing processes~\cite{stefanucci_semiconductor_2024,mocatti_nonequilibrium_2025}. Our approach is similar to the excitonic Bloch equations from ref.~\cite{Stefanucci_ExcitonicBlochEquations_2025} in the low-density regime, except for one key difference: the exciton-phonon coupling  $\gamma$ corresponds to the unscreened coupling strength.

The quantum kinetic equation of motion \eqref{eq:eom_density_matrix}, using the experimentally relevant laser pulse parameters, provides a microscopic description of the exciton dynamics in BP, capturing  the coherent exciton generation and incoherent relaxation on the same footing. To directly compare with the experimental trARPES data, we compute the time-resolved photoemission intensity following refs.~\cite{rustagi_photoemission_2018,beaulieu_berry_2024,Stefanucci_timeresolved_2026,schuler_theory_2021}:
\begin{align}
	\label{eq:trARPES_intensity}
	I(\vec{k}_\parallel,\omega,\tau) = & \sum_{\vec{Q}} N_{\vec{Q}}(\tau) \int dk_z F(k_z - p_\perp) |M_c(\vec{k},p_\perp)|^2 \nonumber \\
	&\times |\Lambda^{\lambda,\vec{Q}}(\vec{k}, k_z)|^2 \delta(\varepsilon_v(\vec{k} - \vec{Q}) + E_x(\vec{Q}) + \omega_\mathrm{pr} - \omega) \nonumber 
    \\ & + I_\text{LAPE}(\vec{k}_\parallel,\omega,\tau) \ .
	% & \delta(\varepsilon_v(\vec{k})) \ .
\end{align}
\FloatBarrier
Here, $\vec{k}_\parallel$ and $\omega$ are the in-plane momentum and energy of the emitted photoelectrons, $\tau$ is the pump-probe delay, and $\omega_\mathrm{pr}$ is the probe photon energy. Going beyond the usual treatment of 2D systems, we also include the out-of-plane momentum $p_\perp$ of the photoelectrons (determined by energy conservation) and account for the finite mean-free path by the Lorentzian function $F(k_z - p_\perp)$. Note that we have specialized to the case of a single exciton band $\lambda$ and single valence/conduction band for clarity. To account for the pronounced polarization dependence of the ARPES signal, we include the photoemission matrix element of the conduction band $M_c(\vec{k},p_\perp)$, which we compute as in ref.~\cite{bao_revealing_2024}.
As the first step we fixed the peak field strength $E^{(0)}$ by computing the time-integrated EDC and matching the intensity of the exciton peak to the experimental data [Fig.~\ref{fig1}(f)].  This yields an excitation density of about $n_e \simeq 2\times 10^{16}$ cm$^{-3}$, which corresponds to the low-density limit for 3D semiconductors with well-defined excitons~\cite{Klingshirn2012}. Before we can compare the time-resolved photoemission intensity, Eq.~\eqref{eq:trARPES_intensity}, to the experimental data, we need to account for LAPE effects. While the experimental geometry has been calibrated to minimize LAPE contributions for $s$-polarized pump pulses, a residual LAPE signal is still observed in the experiments. This is potentially due to imperfect polarization
control in the experimental setup, leading to a small $p$-polarized component in the pump beam. Another possibility is the non-trivial interaction of the pump field with the sample surface -- which has been discussed as Coulomb-laser coupling in the context of attosecond photoemission from atoms, which can induce LAPE even with nominal $s$-polarization. This effect is probably more important in our case due to the low photon energy of the probe. 
In trARPES, the LAPE contribution appears as a replica of the VB shifted by the pump photon energy; its intensity scales with the pump-probe time overlap and decays on the timescale of the pump pulse duration. This is accounted for by the term $I_\text{LAPE}(\vec{k}_\parallel, \omega, \tau)$ in Eq.~\eqref{eq:trARPES_intensity} (see Supplemental Materials for more details).
\\

\textbf{Ultrafast dynamics of excitons}

With these ingredients, we can directly compare the simulated trARPES intensity to the experimental data of Fig.~\ref{fig:trARPES_exp_theory}. In the calculated spectrum at $t_0$ [Fig.~\ref{fig:trARPES_exp_theory}(h)], we can clearly identify the exciton peak below the CB, as well as the faint spectral features dispersing downward. This is the signature expected for coherent excitons ($N_{\vec{Q}} = 0$ for $\vec{Q}\ne 0$)~\cite{rustagi_photoemission_2018}, as the energy conservation condition in Eq.~\eqref{eq:trARPES_intensity} corresponds to the VB dispersion shifted by the exciton energy $E_x(\vec{Q}=0)$. Similar features are present in the experimental data, confirming the excitonic nature of the observed in-gap signal. 
\begin{figure}[t]
\centering
\includegraphics[width=\linewidth]{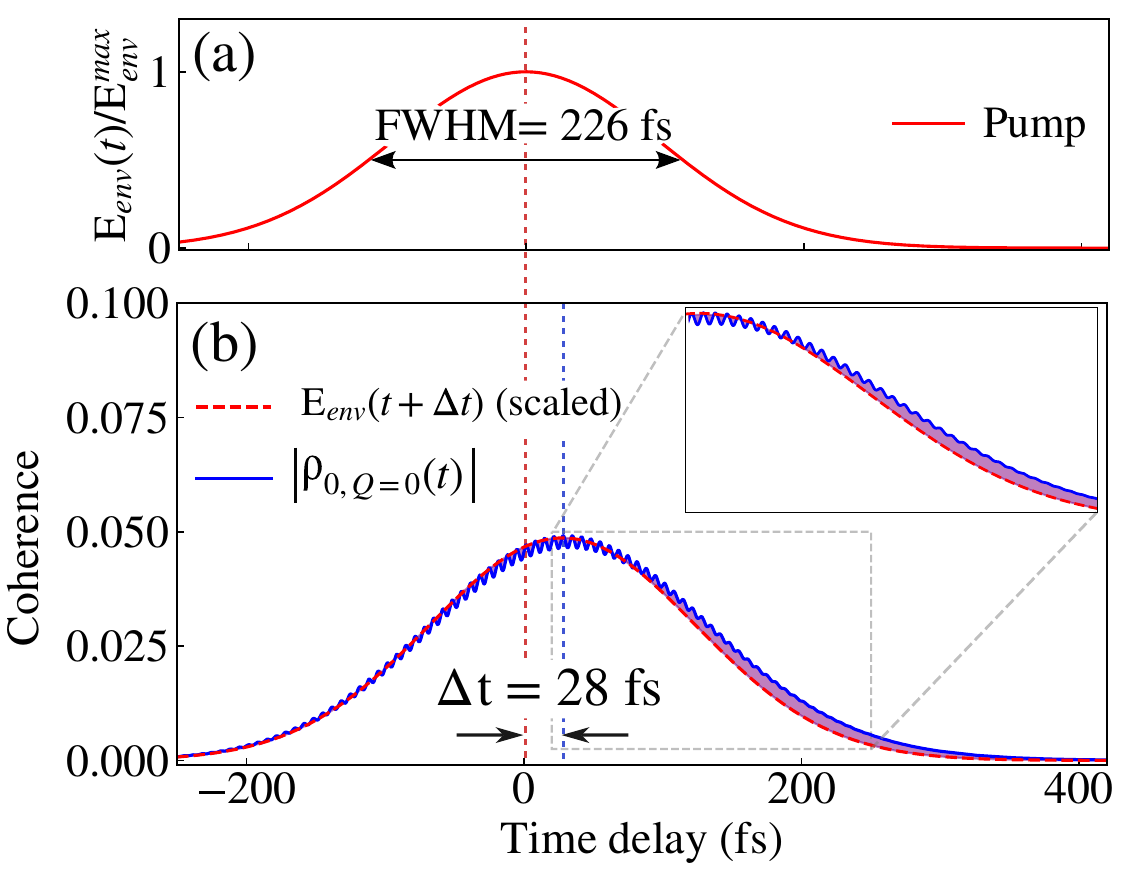}
\caption{\textbf{Exciton coherence.} (a) Envelope of the IR pump pulse and its corresponding FWHM of $226$ fs, normalized to its maximum. (b) Absolute value of the coherence between the ground state and the coherent exciton, $\rho_{0,\mathbf{Q=0}}(t)$. Together with it, we plot the envelope of the electric field scaled to the maximum of the absolute value of the coherence and shifted by $\Delta t$ (time difference between the maximum of the coherence and the pump pulse). Thus, we highlight the shift between them and the asymmetric behavior of the coherence in comparison with the field.
} 
\label{fig:coherences}
\end{figure}
Now we revisit the transient dynamics in the three distinct ROI indicated by the colored boxes in Fig.~\ref{fig:trARPES_exp_theory}(a): (i) the ROI with maximum exciton intensity (green box), (ii) an intermediate ROI capturing the coherent exciton signal (blue box), and (iii) a ROI in the gap (red box). The corresponding transient dynamics extracted from the experimental and simulated trARPES data are shown in Fig.~\ref{fig:trARPES_exp_theory}(b)-(g). 
Achieving agreement between experiment and theory across all ROI is only possible by a unique choice of the effective exciton-phonon coupling strength ($\gamma \simeq 2.5\times 10^{-4}$~a.u.) and LAPE intensity. 
%With our system is within the low-density limit
First-principles derivations of the Excitonic Bloch Equations (EBEs) have shown that standard model Hamiltonians are often overscreened in the electron-phonon couplings (due to a double-counting of the dielectric response). However, by fitting $\gamma$ directly to the experimental trARPES dynamics and staying in the low density regime, our effective $\gamma$ corresponds to the irreducible coupling case, capturing the physical scattering rates \cite{Stefanucci_ExcitonicBlochEquations_2025,Stefanucci_timeresolved_2026}.
The lowest energy ROI in Fig.~\ref{fig:trARPES_exp_theory}(d),(g) shows a pronounced peak close to delay $\tau=0$ that quickly decays. Nevertheless, a slightly asymmetric tail is visible. The theory indicates that this part of the signal is due to LAPE, with a small exciton contribution responsible for the asymmetry. Moving to the intermediate ROI [see Fig.~\ref{fig:trARPES_exp_theory}(c),(f)], the exciton contribution becomes more prominent, leading to a clear asymmetric peak with a delayed maximum and persistent intensity on the picosecond time scale. Finally, in the ROI with maximum exciton intensity [Fig.~\ref{fig:trARPES_exp_theory}(b),(e)], the LAPE contribution is negligible, and the transient dynamics is fully dominated by the exciton population. Here, we observe a pronounced delayed rise of the signal, reflecting the phonon-assisted scattering of excitons from $\vec{Q}=0$ to nonzero momentum states. 
Overall, we find excellent agreement between experiment and theory for all three ROI, confirming our microscopic understanding of the ultrafast exciton dynamics in BP.

The different time delays of the maximum intensity in the three ROI can be understood by the transition from a coherent exciton population at $\vec{Q}=0$ to an incoherent population at $\vec{Q}\ne 0$. Using the predictive power of our quantum-kinetic model, we can directly track the time evolution of the exciton populations $N_{\vec{Q}}(t)$ and coherences $\rho_{0\vec{Q}}(t)$. In Fig.~\ref{fig:exciton_population}(a)-(d), we show the simulated trARPES intensity at different time delays and in Fig.~\ref{fig:exciton_population}(e)-(h) the corresponding exciton populations. As time progresses, the initially coherent exciton population at $\vec{Q}=0$ (bright excitons) rapidly scatters with acoustic phonons, redistributing the exciton population to nonzero momentum states (dark excitons). In the trARPES intensity, this leads to a shift of the maximum intensity towards the CBM, indicated by  the gray horizontal lines in Fig.~\ref{fig:exciton_population}(a)-(d).

In our theory we have direct access to the coherence dynamics $\rho_{0,\vec{Q}=0}(t)$ (corresponding to a macroscopic time-dependent polarization), which is not directly measurable in our experiments. 
The coherence closely follows the shape of the laser pulse [Fig.~\ref{fig:coherences}(a),(b)], indicating that the coherent exciton population is predominantly present during the pump pulse duration and decays rapidly afterwards (indicated by the asymmetric shape). The delay between the peak of the coherence and the pump $\Delta t \sim 28$~fs is related to both population and decoherence dynamics. With the exciton-phonon coupling strength $\gamma$ determined above, we obtain a coherence lifetime from the tail asymmetry, obtaining $T_\mathrm{coh} \simeq 40$~fs.

\begin{figure*}[ht]
\centering
\includegraphics[width=\linewidth]{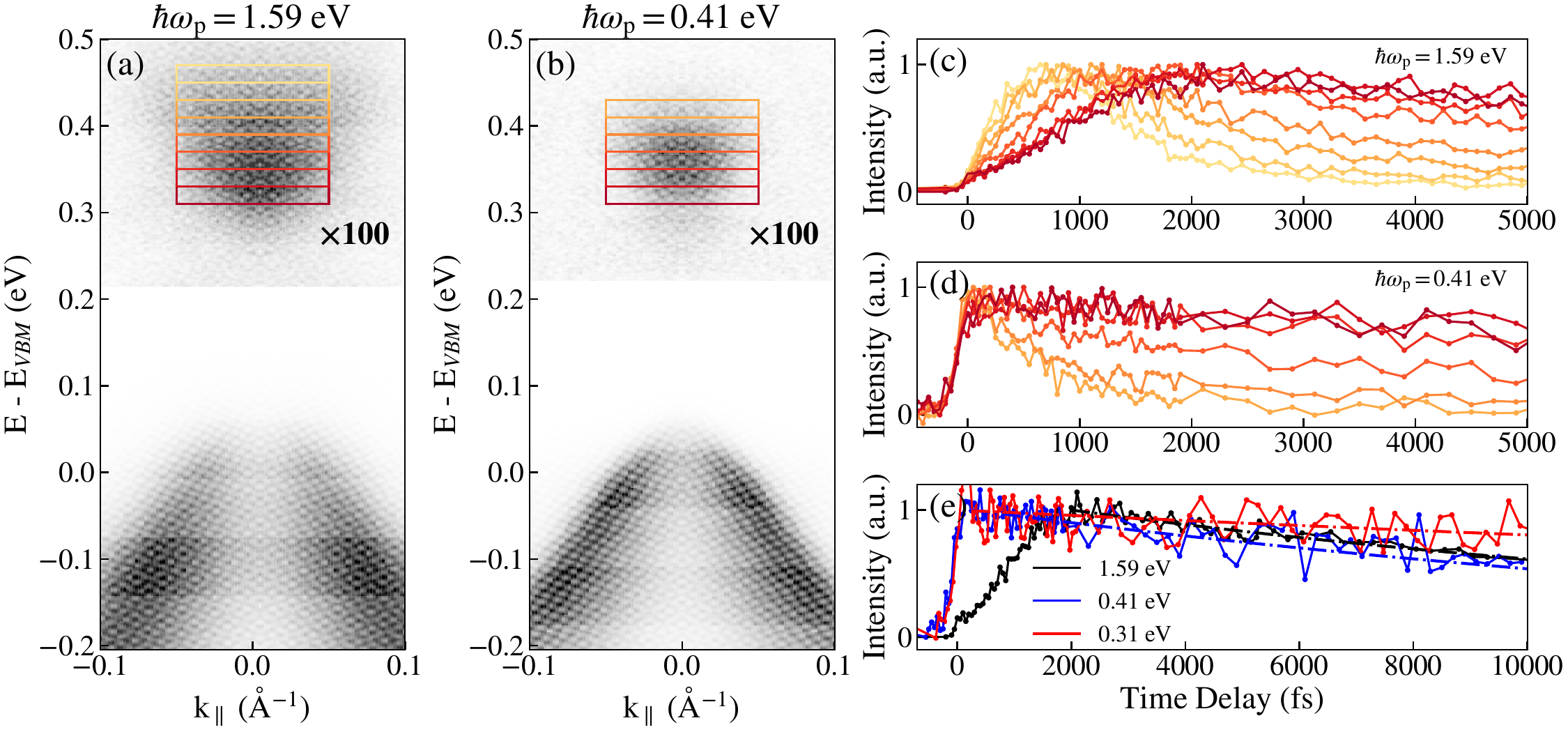}
\caption{\textbf{Comparison of the Dynamics in different Pump Regimes.} (a) Pump-probe ARPES spectra acquired with a pump photon energy of $\hbar\omega_\mathrm{p} = 1.55$~eV integrated from t0 to +1 ps. (b) Pump-probe ARPES spectra acquired with a pump photon energy of $\hbar\omega_\mathrm{p} = 0.41$~eV, integrated from t0 to +1 ps. (c) Normalized time traces extracted in the ROI highlighted in (a). (d) Normalized time traces extracted in the ROI highlighted in (b). (e) Comparison of the long-time relaxation of the signal at the CBM in the lowest energy ROI in (a) and (b). The red curve was extracted at the corresponding energy-momentum ROI in the spectrum with a pump photon energy of $\hbar\omega_\mathrm{p} = 0.31$~eV, shown in Fig.~\ref{fig1}(d). To quantify the interband relaxation, a fit with a single exponential decay function was performed to reveal respective carrier lifetimes of $\tau = 16.5$~ps ($1.59$~eV), $20.6$~ps ($0.41$~eV), and $56.4$~ps ($0.31$~eV). All curves are normalized w.r.t the starting value of the fit.}
\label{fig4}
\end{figure*}

To systematically explore excitation regimes that are off-resonant with respect to the exciton, we performed further trARPES measurements with a pump photon energy of $\hbar\omega_\mathrm{p} = 1.59$~eV. Within this regime, multiple studies on the non-equilibrium dynamics have been conducted. Previous trARPES works revealed that photoexcited BP hosts a variety of non-equilibrium effects, including surface photovoltage~\cite{kremer_ultrafast_2021}, transient Stark broadening and band gap renormalization at high excitation densities~\cite{chen_band_2019,hedayat_non-equilibrium_2021,roth_photocarrier-induced_2019}. 
Here, we want to focus on the population dynamics at $\hbar\omega_\mathrm{p} = 1.59$~eV in comparison to data taken at smaller pump photon energies discussed in relation to Fig.~\ref{fig1}.
The trARPES spectrum in Fig.~\ref{fig4}(a) reveals that under this photon excitation energy, electrons are injected in states far above the CBM, partially exceeding the detector window. To highlight the intraband dynamics, we choose ROI in ascending energy and trace the respective signal intensity as a function of delay time. In Fig.~\ref{fig4}(c), we resolve a clear cascade of excited carriers from the highest ROI in energy to the CBM. Remarkably, the relaxation dynamics proceed on an unexpectedly slow timescale: the signal in the highest energy ROI (yellow box) only reaches its maximum 750 fs after excitation, and the population at the CBM peaks after nearly 2 ps. 
In contrast, the excitation with photon energy just above the band gap ($\hbar\omega_\mathrm{p} = 0.41$~eV) directly populates the CBM at $\bar{\Gamma}$. Consequently, only the low-energy region of the CB is occupied, as shown in Fig.~\ref{fig4}(b). The direct excitation into the CB also manifests in the intraband relaxation in Fig. \ref{fig4}(d). The regions of higher energy reach their maximum signal almost immediately after the initial excitation and even in the lowest energy region, the signal peaks after 100 fs only, highlighting the much faster dynamics.
This difference in the time scales suggests that the relaxation pathway strongly depends on the energy of the photoexcited carriers. At high pump photon energies, higher energy states in the CB can be directly populated. The transition from these high-energy states to lower energy states in the CB is emerging as a potential bottleneck for relaxation~\cite{shen_ultrafast_2025}. 
\\

\textbf{Charge carrier dynamics above the band gap}

To study the recombination of photoexcited electrons and holes, in Fig.~\ref{fig4}(e) we compare the dynamics of carriers at the bottom of the CB for the three different excitation regimes, corresponding to the formation of excitons (red), the direct occupation of the CB (blue), and the indirect filling of the CB (black). To isolate the recombination dynamics from intraband relaxation mechanisms, only the signals in the lowest energy ROI (dark red box in Fig.~\ref{fig4}(a) and Fig.~\ref{fig4}(b) and the corresponding position in Fig.~\ref{fig1}(d)) are shown.
Despite the pronounced difference in the rise time, the curves show a similar decay of free carriers in the CB, regardless of whether they were excited with the 1.59 eV or the 0.41 eV pulses. Exponential fits support this, yielding relaxation times of 16.5 ps and 20.6 ps and indicating that the recombination dynamics are nearly independent of the initial excitation and relaxation pathway.
For the case of a photoexcitation resonant with excitonic states (0.31 eV), the relaxation time is much longer, namely 56.4 ps. This is consistent with the fact that for excitons, being bound electron-hole pairs, additional energy must be provided to break the pairs before recombination can take place, enhancing their relaxation time in comparison to free charge carriers. This also indicates that a negligible amount of free carriers are excited at this pump photon energy.

\section*{Discussion and conclusion}

In this study, we generated coherent excitons with zero center-of-mass momentum by direct photoexcitation with a photon energy similar to the band gap of BP. Experimentally, evidence for the creation of such excitons comes from the energy position of the excitonic peak in comparison to free carriers excited in the CB with off-resonant photoexcitation. Additionally, we followed their transient dynamics over a few picoseconds and observed that the in-gap signal due to the excitonic contribution displays a distinct energy dependence. Moreover, by carefully extracting the relaxation time of free charge carriers excited at higher energies, we singled out the relaxation time of the excitons that extends to longer delay times.

A quantum-kinetic calculation based on the Bethe-Salpeter equation and extended to the semiconductor Bloch equation combined with the time-dependent Boltzmann equation including coupling to acoustic phonons allowed us to interpret the experimental data with high fidelity. In about 30 fs, scattering with phonon leads to the full decoherence of the excitons which are scattered into dark excitons carrying non-zero center-of-mass momentum. 
From the comparison to experimental curves, we extracted a quantitative value for the exciton-phonon coupling strength, which sets a benchmark for state-of-the-art \emph{ab intio} methods for calculating exciton-phonon coupling and exciton linewidth~\cite{sangalli_ab-initio_2018,chan_exciton_2023, chan_excitonphonon_2024}.
The scattering process redistributing the excitons in their intrinsic dispersion is driven by phonon absorption. This is at odds with other recent experiments probing excitons that are generated indirectly by a non-resonant process, after which  excited charge carriers relax towards the band extrema \cite{Gosetti2025,Trovatello2020}. In these experiments, the excited free carriers, and eventually the excitons, emit phonons, leading to the warming of the lattice.

Our work reveals that exciton scattering with acoustic phonons is highly efficient in three‑dimensional bulk BP, leading to a rapid loss of excitonic coherence on ultrashort timescales. Importantly, we identify intravalley phonon scattering as the sole decoherence mechanism in this system, in stark contrast to transition‑metal dichalcogenides, where intervalley scattering into momentum‑forbidden states is widely recognized as the dominant relaxation pathway \cite{Fanciulli2023,Madeo2020,Bertoni2016}. This result demonstrates that excitonic coherence can be fundamentally limited by intravalley processes alone, even in the absence of a multivalley band structure. As a consequence, our findings place stringent constraints on the possibility of coherently manipulating excitons — and thereby the electronic structure of bulk BP — using optical excitation beyond the initial ultrafast regime. Extending the combined experimental and theoretical approach presented here to few‑layer BP, where reduced dimensionality and modified phonon spectra may alter the exciton–phonon coupling, represents an exciting direction for future investigations.

\section*{Materials and methods}

\subsection*{Sample growth}

Large single crystals of BP were synthesized by a modified chemical vapor transport technique as reported in Ref.~\cite{Wang2019}. Red phosphorus (400 mg, $> 99.7\%$, powder), SnI$_4$ (8 mg, 99\%), and Sn (16 mg, 99.995\%, powder) were sealed in an evacuated quartz ampoule. The ampoule was placed horizontally in a muffle furnace and partially covered on one side with ceramic elements in order to establish a controlled temperature gradient along the ampoule. 
The furnace was heated to 650$^\circ$C at a rate of 600$^\circ$C per hour, followed by a multi-step cooling program consisting of cooling to 550$^\circ$C over 1\,h, to 500$^\circ$C over 8\,h, and subsequently to 200$^\circ$C over 2\,h, followed by rapid cooling by switching off the furnace. 
Manually selected crystals were sealed in an evacuated quartz ampoule and placed in a tubular furnace with one end of the ampoule extending outside the heated zone. The sample was heated to 200$^\circ$C for 3\,h in order to remove residual SnI$_4$ and elemental iodine. The obtained crystals were then rinsed sequentially with acetone (Ac), water, and analytical ethanol (EtOH), followed by drying under dynamic vacuum.

\subsection*{Time-resolved ARPES}

trARPES experiments were carried out using a Scienta DA30 hemispherical photoelectron analyzer with a base pressure better than $5 \times 10^{-11}$ mbar. For the light source, half of the power of a femtosecond laser (Pharos, Light Conversion, operating at 1030 nm) is converted into 780 nm (1.59 eV) light with an optical parametric amplifier, which is then frequency-quadrupled to 6.3 eV in $\beta-$BaB$_2$O$_4$ crystals to generate UV pulses (see e.g. ref.~\cite{faure_full_2012}), which were later-on $p$-polarized. The intrinsic resolution of the UV pulse is 25 meV, as determined by the fit of the Fermi edge of a polycrystalline metal. The remaining half of the fundamental laser power is directed into a collinear optical parametric amplifier (Orpheus, Light Conversion) to generate IR pulses at 1.55, 0.41 or 0.31 eV. The total temporal resolution was determined to be better than 100 fs, 160 fs and 240 fs, respectively, by measuring the width of the photoemission cross-correlation between the pump and the probe pulses. The pump-probe measurements were performed at 200 kHz and the incident fluence used to detect the exciton was of the order of $F\approx 46$ $\mu$J/cm$^2$. Unless specified otherwise, the pump laser was $s$-polarized.

The slit of our hemispherical analyzer is oriented such that the parallel momentum is running along the armchair direction. All data have been acquired at the temperature of  $T_0 = 80$~K.

\bibliography{main_refs,Supplemental_material/sup_refs}

@article{Cappellini1993,
  title = {Model dielectric function for semiconductors},
  author = {Cappellini, G. and Del Sole, R. and Reining, Lucia and Bechstedt, F.},
  journal = {Phys. Rev. B},
  volume = {47},
  issue = {15},
  pages = {9892--9895},
  numpages = {0},
  year = {1993},
  month = {Apr},
  publisher = {American Physical Society},
  doi = {10.1103/PhysRevB.47.9892},
  url = {https://link.aps.org/doi/10.1103/PhysRevB.47.9892}
}

@article{Zhu2014,
  title = {Coexistence of size-dependent and size-independent thermal conductivities in phosphorene},
  author = {Zhu, Liyan and Zhang, Gang and Li, Baowen},
  journal = {Phys. Rev. B},
  volume = {90},
  issue = {21},
  pages = {214302},
  numpages = {6},
  year = {2014},
  month = {Dec},
  publisher = {American Physical Society},
  doi = {10.1103/PhysRevB.90.214302},
  url = {https://link.aps.org/doi/10.1103/PhysRevB.90.214302}
}

@article{Yates2007,
  title = {{Spectral and Fermi surface properties from Wannier interpolation}},
  author = {Yates, Jonathan R. and Wang, Xinjie and Vanderbilt, David and Souza, Ivo},
  journal = {Phys. Rev. B},
  volume = {75},
  issue = {19},
  pages = {195121},
  numpages = {11},
  year = {2007},
  month = {May},
  publisher = {American Physical Society},
  doi = {10.1103/PhysRevB.75.195121},
  url = {https://link.aps.org/doi/10.1103/PhysRevB.75.195121}
}

@article{Hartmut2001,
    doi = {10.1088/0953-8984/13/34/313},
    url = {https://doi.org/10.1088/0953-8984/13/34/313},
    year = {2001},
    month = {aug},
    publisher = {},
    volume = {13},
    number = {34},
    pages = {7679},
    author = {Hartmut Zabel},
    title = {Phonons in layered compounds},
    journal = {Journal of Physics: Condensed Matter},
}

@article{Politano2012,
    title = {Elastic properties of a macroscopic graphene sample from phonon dispersion measurements},
    journal = {Carbon},
    volume = {50},
    number = {13},
    pages = {4903-4910},
    year = {2012},
    issn = {0008-6223},
    doi = {https://doi.org/10.1016/j.carbon.2012.06.019},
    url = {https://www.sciencedirect.com/science/article/pii/S0008622312005258},
    author = {Antonio Politano and Antonio Raimondo Marino and Davide Campi and Daniel Farías and Rodolfo Miranda and Gennaro Chiarello},
}

@article{Fujii1982,
    title = {Inelastic neutron scattering study of acoustic phonons of black phosphorus},
    journal = {Solid State Communications},
    volume = {44},
    number = {5},
    pages = {579-582},
    year = {1982},
    issn = {0038-1098},
    doi = {https://doi.org/10.1016/0038-1098(82)90558-0},
    url = {https://www.sciencedirect.com/science/article/pii/0038109882905580},
    author = {Y. Fujii and Y. Akahama and S. Endo and S. Narita and Y. Yamada and G. Shirane},
}

@Article{Guangzhao2015,
    author ="Qin, Guangzhao and Yan, Qing-Bo and Qin, Zhenzhen and Yue, Sheng-Ying and Hu, Ming and Su, Gang",
    title  ="Anisotropic intrinsic lattice thermal conductivity of phosphorene from first principles",
    journal  ="Phys. Chem. Chem. Phys.",
    year  ="2015",
    volume  ="17",
    issue  ="7",
    pages  ="4854-4858",
    publisher  ="The Royal Society of Chemistry",
    doi  ="10.1039/C4CP04858J",
    url  ="http://dx.doi.org/10.1039/C4CP04858J",
}

@article{Li2014,
    title = {ShengBTE: A solver of the {Boltzmann} transport equation for phonons},
    journal = {Computer Physics Communications},
    volume = {185},
    number = {6},
    pages = {1747-1758},
    year = {2014},
    issn = {0010-4655},
    doi = {https://doi.org/10.1016/j.cpc.2014.02.015},
    url = {https://www.sciencedirect.com/science/article/pii/S0010465514000484},
    author = {Wu Li and Jesús Carrete and Nebil {A. Katcho} and Natalio Mingo},
}

@article{mahmood_selective_2016,
	title = {Selective scattering between {Floquet}–{Bloch} and {Volkov} states in a topological insulator},
	volume = {12},
	copyright = {2016 Springer Nature Limited},
	issn = {1745-2481},
	url = {https://www.nature.com/articles/nphys3609},
	doi = {10.1038/nphys3609},
	number = {4},
	urldate = {2026-03-25},
	journal = {Nature Physics},
	publisher = {Nature Publishing Group},
	author = {Mahmood, Fahad and Chan, Ching-Kit and Alpichshev, Zhanybek and Gardner, Dillon and Lee, Young and Lee, Patrick A. and Gedik, Nuh},
	month = apr,
	year = {2016},
	pages = {306--310},
}

@article{saathoff_laser-assisted_2008,
	title = {Laser-assisted photoemission from surfaces},
	volume = {77},
	url = {https://link.aps.org/doi/10.1103/PhysRevA.77.022903},
	doi = {10.1103/PhysRevA.77.022903},
	number = {2},
	urldate = {2026-03-25},
	journal = {Physical Review A},
	publisher = {American Physical Society},
	author = {Saathoff, G. and Miaja-Avila, L. and Aeschlimann, M. and Murnane, M. M. and Kapteyn, H. C.},
	month = feb,
	year = {2008},
	pages = {022903},

}

@software{wortmann_2023_7778444,
  author       = {Wortmann, Daniel and
                  Michalicek, Gregor and
                  Hilgers, Robin and
                  Neukirchen, Alexander and
                  Janssen, Henning and
                  Grytsiuk, Uliana and
                  Broeder, Jens and
                  Gerhorst, Christian-Roman},
  title        = {FLEUR},
  month        = mar,
  year         = 2023,
  publisher    = {Zenodo},
  version      = {testtag},
  doi          = {10.5281/zenodo.7778444},
  url          = {https://doi.org/10.5281/zenodo.7778444},
}

@article{Aryasetiawan2004,
  title = {Frequency-dependent local interactions and low-energy effective models from electronic structure calculations},
  author = {Aryasetiawan, F. and Imada, M. and Georges, A. and Kotliar, G. and Biermann, S. and Lichtenstein, A. I.},
  journal = {Phys. Rev. B},
  volume = {70},
  issue = {19},
  pages = {195104},
  numpages = {8},
  year = {2004},
  month = {Nov},
  publisher = {American Physical Society},
  doi = {10.1103/PhysRevB.70.195104},
  url = {https://link.aps.org/doi/10.1103/PhysRevB.70.195104}
}

@article{LAPE_2006,
  title = {{Laser-Assisted Photoelectric Effect from Surfaces}},
  author = {Miaja-Avila, L. and Lei, C. and Aeschlimann, M. and Gland, J. L. and Murnane, M. M. and Kapteyn, H. C. and Saathoff, G.},
  journal = {Phys. Rev. Lett.},
  volume = {97},
  issue = {11},
  pages = {113604},
  numpages = {4},
  year = {2006},
  month = {Sep},
  publisher = {American Physical Society},
  doi = {10.1103/PhysRevLett.97.113604},
  url = {https://link.aps.org/doi/10.1103/PhysRevLett.97.113604}
}

@article{Schueler2022,
  author   = {Schüler, Michael and Beaulieu, Samuel},
  title    = {Probing topological {Floquet} states in {{WSe}}{\textsubscript{2}} using circular dichroism in time- and angle-resolved photoemission spectroscopy},
  journal  = {Communications Physics},
  year     = {2022},
  month    = {Jun},
  day      = {27},
  volume   = {5},
  number   = {1},
  pages    = {164},
  doi      = {10.1038/s42005-022-00944-w},
  url      = {https://doi.org/10.1038/s42005-022-00944-w},
  issn     = {2399-3650}
}

@misc{mittenzwey_excitonic_2025,
      title={Excitonic Theory of the Ultrafast Optical Response of 2D-Quantum-Confined Semiconductors at Elevated Densities}, 
      author={Henry Mittenzwey and Oliver Voigt and Andreas Knorr},
      year={2026},
      eprint={2512.03198},
      archivePrefix={arXiv},
      primaryClass={cond-mat.mes-hall},
      url={https://arxiv.org/abs/2512.03198}, 
}

@article{dogadov_dissecting_2026,
   title       = {Dissecting intervalley coupling mechanisms in monolayer transition metal dichalcogenides},
   volume      = {10},
   url         = {https://www.nature.com/articles/s41699-025-00653-2},
   doi         = {10.1038/s41699-025-00653-2},
   number      = {1},
   journal     = {npj 2D Mater Appl},
   author      = {Dogadov, Oleg and Mittenzwey, Henry and Bertolotti, Micol and Olsen, Nicholas and Deckert, Thomas and Trovatello, Chiara and Zhu, Xiaoyang and Brida, Daniele and Cerullo, Giulio and Knorr, Andreas and Dal Conte, Stefano},
   year        = {2026},
   pages       = {21},
}

@article{schuler_theory_2021,
   title       = {Theory of subcycle time-resolved photoemission: {Application} to terahertz photodressing in graphene},
   volume      = {253},
   copyright   = {All rights reserved},
   url         = {https://www.sciencedirect.com/science/article/pii/S0368204821000736},
   doi         = {10.1016/j.elspec.2021.147121},
   journal     = {Journal of Electron Spectroscopy and Related Phenomena},
   author      = {Sch\"{u}ler, Michael and Sentef, Michael A.},
   year        = {2021},
   pages       = {147121},
}

@article{chen_first-principles_2022,
   title       = {First-principles ultrafast exciton dynamics and time-domain spectroscopies: {Dark}-exciton mediated valley depolarization in monolayer {{WSe}}{\textsubscript{2}}},
   volume      = {4},
   url         = {https://link.aps.org/doi/10.1103/PhysRevResearch.4.043203},
   doi         = {10.1103/PhysRevResearch.4.043203},
   number      = {4},
   journal     = {Phys. Rev. Research},
   author      = {Chen, Hsiao-Yi and Sangalli, Davide and Bernardi, Marco},
   year        = {2022},
   pages       = {043203},
}

@article{lechifflart_first-principles_2023,
   title       = {First-principles study of luminescence in hexagonal boron nitride single layer: {Exciton}-phonon coupling and the role of substrate},
   volume      = {7},
   url         = {https://link.aps.org/doi/10.1103/PhysRevMaterials.7.024006},
   doi         = {10.1103/PhysRevMaterials.7.024006},
   number      = {2},
   journal     = {Phys. Rev. Materials},
   author      = {Lechifflart, Pierre and Paleari, Fulvio and Sangalli, Davide and Attaccalite, Claudio},
   year        = {2023},
   pages       = {024006},
}

@article{kasprzak_boseeinstein_2006,
   title       = {Bose-{Einstein} condensation of exciton polaritons},
   volume      = {443},
   copyright   = {http://www.springer.com/tdm},
   url         = {https://www.nature.com/articles/nature05131},
   doi         = {10.1038/nature05131},
   number      = {7110},
   journal     = {Nature},
   author      = {Kasprzak, J. and Richard, M. and Kundermann, S. and Baas, A. and Jeambrun, P. and Keeling, J. M. J. and Marchetti, F. M. and Szymańska, M. H. and Andr\'{e}, R. and Staehli, J. L. and Savona, V. and Littlewood, P. B. and Deveaud, B. and Dang, Le Si},
   year        = {2006},
   pages       = {409--414},
}

@article{chan_exciton_2025,
   title       = {Exciton thermalization dynamics in monolayer {{MoS}}{\textsubscript{2}} : {A} first-principles {Boltzmann} equation study},
   volume      = {111},
   url         = {https://link.aps.org/doi/10.1103/PhysRevB.111.184305},
   doi         = {10.1103/PhysRevB.111.184305},
   number      = {18},
   journal     = {Phys. Rev. B},
   author      = {Chan, Yang-hao and Haber, Jonah B. and Naik, Mit H. and Louie, Steven G. and Neaton, Jeffrey B. and Da Jornada, Felipe H. and Qiu, Diana Y.},
   year        = {2025},
   pages       = {184305},
}

@article{gucci_encoding_2026,
   title       = {Encoding and manipulating ultrafast coherent valleytronic information with lightwaves},
   volume      = {20},
   url         = {https://www.nature.com/articles/s41566-025-01823-w},
   doi         = {10.1038/s41566-025-01823-w},
   number      = {3},
   journal     = {Nat. Photon.},
   author      = {Gucci, Francesco and Molinero, Eduardo B. and Russo, Mattia and San-Jose, Pablo and Camargo, Franco V. A. and Maiuri, Margherita and Ivanov, Misha and Jim\'{e}nez-Gal\'{a}n, \'{A}lvaro and Silva, Rui E. F. and Dal Conte, Stefano and Cerullo, Giulio},
   year        = {2026},
   pages       = {266--272},
}

@article{Wang2019,
  author  = {Wang, Dongya and Yi, Peng and Wang, Lin and Zhang, Lu and
             Li, Hai and Lu, Min and Xie, Xiaoji and Huang, Ling and
             Huang, Wei},
  title   = {{Revisiting the Growth of Black Phosphorus in Sn-I Assisted Reactions}},
  journal = {Frontiers in Chemistry},
  volume  = {7},
  pages   = {21},
  year    = {2019},
  doi     = {10.3389/fchem.2019.00021},
  url     = {https://www.frontiersin.org/journals/chemistry/articles/10.3389/fchem.2019.00021},
  issn    = {2296-2646}
}

@article{Pizzi2020,
    doi = {10.1088/1361-648X/ab51ff},
    url = {https://doi.org/10.1088/1361-648X/ab51ff},
    year = {2020},
    month = {jan},
    publisher = {IOP Publishing},
    volume = {32},
    number = {16},
    pages = {165902},
    author = {Pizzi, Giovanni and Vitale, Valerio and Arita, Ryotaro and Blügel, Stefan and Freimuth, Frank and Géranton, Guillaume and Gibertini, Marco and Gresch, Dominik and Johnson, Charles and Koretsune, Takashi and Ibañez-Azpiroz, Julen and Lee, Hyungjun and Lihm, Jae-Mo and Marchand, Daniel and Marrazzo, Antimo and Mokrousov, Yuriy and Mustafa, Jamal I and Nohara, Yoshiro and Nomura, Yusuke and Paulatto, Lorenzo and Poncé, Samuel and Ponweiser, Thomas and Qiao, Junfeng and Thöle, Florian and Tsirkin, Stepan S and Wierzbowska, Małgorzata and Marzari, Nicola and Vanderbilt, David and Souza, Ivo and Mostofi, Arash A and Yates, Jonathan R},
    title = {Wannier90 as a community code: new features and applications},
    journal = {Journal of Physics: Condensed Matter},
}

@article{Chen2022,
  title = {First-principles ultrafast exciton dynamics and time-domain spectroscopies: Dark-exciton mediated valley depolarization in monolayer {{WSe}}{\textsubscript{2}}},
  author = {Chen, Hsiao-Yi and Sangalli, Davide and Bernardi, Marco},
  journal = {Phys. Rev. Res.},
  volume = {4},
  issue = {4},
  pages = {043203},
  numpages = {10},
  year = {2022},
  month = {Dec},
  publisher = {American Physical Society},
  doi = {10.1103/PhysRevResearch.4.043203},
  url = {https://link.aps.org/doi/10.1103/PhysRevResearch.4.043203}
}

@article{Helmrich2021,
  title = {Phonon-Assisted Intervalley Scattering Determines Ultrafast Exciton Dynamics in {{MoSe}}{\textsubscript{2}} Bilayers},
  author = {Helmrich, Sophia and Sampson, Kevin and Huang, Di and Selig, Malte and Hao, Kai and Tran, Kha and Achstein, Alexander and Young, Carter and Knorr, Andreas and Malic, Ermin and Woggon, Ulrike and Owschimikow, Nina and Li, Xiaoqin},
  journal = {Phys. Rev. Lett.},
  volume = {127},
  issue = {15},
  pages = {157403},
  numpages = {7},
  year = {2021},
  month = {Oct},
  publisher = {American Physical Society},
  doi = {10.1103/PhysRevLett.127.157403},
  url = {https://link.aps.org/doi/10.1103/PhysRevLett.127.157403}
}

@Article{Wallauer2021,
author={Wallauer, Robert
and Perea-Causin, Raul
and M{\"u}nster, Lasse
and Zajusch, Sarah
and Brem, Samuel
and G{\"u}dde, Jens
and Tanimura, Katsumi
and Lin, Kai-Qiang
and Huber, Rupert
and Malic, Ermin
and H{\"o}fer, Ulrich},
title={Momentum-Resolved Observation of Exciton Formation Dynamics in Monolayer {{WS}}{\textsubscript{2}}},
journal={Nano Letters},
year={2021},
month={Jul},
day={14},
publisher={American Chemical Society},
volume={21},
number={13},
pages={5867-5873},
issn={1530-6984},
doi={10.1021/acs.nanolett.1c01839},
url={https://doi.org/10.1021/acs.nanolett.1c01839}
}

@article{Christiansen2019,
  title = {Theory of exciton dynamics in time-resolved {ARPES}: Intra- and intervalley scattering in two-dimensional semiconductors},
  author = {Christiansen, Dominik and Selig, Malte and Malic, Ermin and Ernstorfer, Ralph and Knorr, Andreas},
  journal = {Phys. Rev. B},
  volume = {100},
  issue = {20},
  pages = {205401},
  numpages = {9},
  year = {2019},
  month = {Nov},
  publisher = {American Physical Society},
  doi = {10.1103/PhysRevB.100.205401},
  url = {https://link.aps.org/doi/10.1103/PhysRevB.100.205401}
}

@Article{Schmitt2022,
author={Schmitt, David
and Bange, Jan Philipp
and Bennecke, Wiebke
and AlMutairi, AbdulAziz
and Meneghini, Giuseppe
and Watanabe, Kenji
and Taniguchi, Takashi
and Steil, Daniel
and Luke, D. Russell
and Weitz, R. Thomas
and Steil, Sabine
and Jansen, G. S. Matthijs
and Brem, Samuel
and Malic, Ermin
and Hofmann, Stephan
and Reutzel, Marcel
and Mathias, Stefan},
title={Formation of moir{\'e} interlayer excitons in space and time},
journal={Nature},
year={2022},
month={Aug},
day={01},
volume={608},
number={7923},
pages={499-503},
issn={1476-4687},
doi={10.1038/s41586-022-04977-7},
url={https://doi.org/10.1038/s41586-022-04977-7}
}

@article{
Chan2023,
author = {Y.-H. Chan  and Diana Y. Qiu  and Felipe H. da Jornada  and Steven G. Louie },
title = {Giant self-driven exciton-{Floquet} signatures in time-resolved photoemission spectroscopy of {{MoS}}{\textsubscript{2}} from time-dependent {GW} approach},
journal = {Proceedings of the National Academy of Sciences},
volume = {120},
number = {32},
pages = {e2301957120},
year = {2023},
doi = {10.1073/pnas.2301957120},
}

@Article{Pareek2026,
author={Pareek, Vivek
and Bacon, David R.
and Zhu, Xing
and Chan, Yang-Hao
and Bussolotti, Fabio
and Menezes, Marcos G.
and Chan, Nicholas S.
and Urquizo, Joel P{\'e}rez
and Watanabe, Kenji
and Taniguchi, Takashi
and Perfetto, Enrico
and Man, Michael K. L.
and Mad{\'e}o, Julien
and Stefanucci, Gianluca
and Qiu, Diana Y.
and Goh, Kuan Eng Johnson
and da Jornada, Felipe H.
and Dani, Keshav M.},
title={Driving {Floquet} physics with excitonic fields},
journal={Nature Physics},
year={2026},
month={Feb},
day={01},
volume={22},
number={2},
pages={209-217},
issn={1745-2481},
doi={10.1038/s41567-025-03132-z},
url={https://doi.org/10.1038/s41567-025-03132-z}
}

@article{Hedin1965,
  title = {{New Method for Calculating the One-Particle Green's Function with Application to the Electron-Gas Problem}},
  author = {Hedin, Lars},
  journal = {Phys. Rev.},
  volume = {139},
  issue = {3A},
  pages = {A796--A823},
  numpages = {0},
  year = {1965},
  month = {Aug},
  publisher = {American Physical Society},
  doi = {10.1103/PhysRev.139.A796},
  url = {https://link.aps.org/doi/10.1103/PhysRev.139.A796}
}

@Article{Kobayashi2023,
author={Kobayashi, Yuki
and Heide, Christian
and Johnson, Amalya C.
and Tiwari, Vishal
and Liu, Fang
and Reis, David A.
and Heinz, Tony F.
and Ghimire, Shambhu},
title={Floquet engineering of strongly driven excitons in monolayer tungsten disulfide},
journal={Nature Physics},
year={2023},
month={Feb},
day={01},
volume={19},
number={2},
pages={171-176},
issn={1745-2481},
doi={10.1038/s41567-022-01849-9},
url={https://doi.org/10.1038/s41567-022-01849-9}
}

@article{Friedrich2010,
  title = {Efficient implementation of the {GW} approximation within the all-electron {FLAPW method}},
  author = {Friedrich, Christoph and Bl\"ugel, Stefan and Schindlmayr, Arno},
  journal = {Phys. Rev. B},
  volume = {81},
  issue = {12},
  pages = {125102},
  numpages = {16},
  year = {2010},
  month = {Mar},
  publisher = {American Physical Society},
  doi = {10.1103/PhysRevB.81.125102},
  url = {https://link.aps.org/doi/10.1103/PhysRevB.81.125102}
}

@article{beaulieu_berry_2024,
   title       = {Berry curvature signatures in chiroptical excitonic transitions},
   volume      = {10},
   url         = {https://www.science.org/doi/full/10.1126/sciadv.adk3897},
   doi         = {10.1126/sciadv.adk3897},
   number      = {26},
   journal     = {Science Advances},
   publisher   = {American Association for the Advancement of Science},
   author      = {Beaulieu, Samuel and Dong, Shuo and Christiansson, Viktor and Werner, Philipp and Pincelli, Tommaso and Ziegler, Jonas D. and Taniguchi, Takashi and Watanabe, Kenji and Chernikov, Alexey and Wolf, Martin and Rettig, Laurenz and Ernstorfer, Ralph and Sch\"{u}ler, Michael},
   year        = {2024},
   pages       = {eadk3897},
}

@misc{Stefanucci_timeresolved_2026,
      title={Unified First-Principles Formula for Time-Resolved ARPES Spectra of Coherent and Incoherent Excitons}, 
      author={Gianluca Stefanucci and Enrico Perfetto},
      year={2026},
      eprint={2601.16786},
      archivePrefix={arXiv},
      primaryClass={cond-mat.mtrl-sci},
      url={https://arxiv.org/abs/2601.16786}, 
}

@article{stefanucci_semiconductor_2024,
   title       = {Semiconductor electron-phonon equations: {A} rung above {Boltzmann} in the many-body ladder},
   volume      = {16},
   url         = {https://scipost.org/10.21468/SciPostPhys.16.3.073},
   doi         = {10.21468/SciPostPhys.16.3.073},
   number      = {3},
   journal     = {SciPost Phys.},
   author      = {Stefanucci, Gianluca and Perfetto, Enrico},
   year        = {2024},
   pages       = {073},
}

@misc{mocatti_nonequilibrium_2025,
   title       = {Nonequilibrium {Photocarrier} and {Phonon} {Dynamics} from {First} {Principles}: a {Unified} {Treatment} of {Carrier}-{Carrier}, {Carrier}-{Phonon}, and {Phonon}-{Phonon} {Scattering}},
   copyright   = {Creative Commons Attribution 4.0 International},
   shorttitle  = {Nonequilibrium {Photocarrier} and {Phonon} {Dynamics} from {First} {Principles}},
   url         = {https://arxiv.org/abs/2512.08618},
   urldate     = {2026-01-20},
   publisher   = {arXiv},
   author      = {Mocatti, Stefano and Marini, Giovanni and Volpato, Giulio and Cudazzo, Pierluigi and Calandra, Matteo},
   year        = {2025},
}

@article{perfetto_real-time_2023,
   title       = {Real-{Time} \textit{{GW}} -{Ehrenfest}-{Fan}-{Migdal} {Method} for {Nonequilibrium} {2D} {Materials}},
   volume      = {23},
   url         = {https://pubs.acs.org/doi/10.1021/acs.nanolett.3c01772},
   doi         = {10.1021/acs.nanolett.3c01772},
   number      = {15},
   journal     = {Nano Lett.},
   author      = {Perfetto, Enrico and Stefanucci, Gianluca},
   year        = {2023},
   pages       = {7029--7036},
}

@article{bao_revealing_2024,
	title = {Manipulating the symmetry of photon-dressed electronic states},
	volume = {15},
	copyright = {2024 The Author(s)},
	issn = {2041-1723},
	url = {https://www.nature.com/articles/s41467-024-54760-7},
	doi = {10.1038/s41467-024-54760-7},
	number = {1},
	urldate = {2026-05-26},
	journal = {Nature Communications},
	publisher = {Nature Publishing Group},
	author = {Bao, Changhua and Schüler, Michael and Xiao, Teng and Wang, Fei and Zhong, Haoyuan and Lin, Tianyun and Cai, Xuanxi and Sheng, Tianshuang and Tang, Xiao and Zhang, Hongyun and Yu, Pu and Sun, Zhiyuan and Duan, Wenhui and Zhou, Shuyun},
	month = dec,
	year = {2024},
	pages = {10535},
}

@article{Bertoni2016,
  title = {Generation and Evolution of Spin-, Valley-, and Layer-Polarized Excited Carriers in Inversion-Symmetric {{WSe}}{\textsubscript{2}}},
  author = {Bertoni, R. and Nicholson, C. W. and Waldecker, L. and H\"ubener, H. and Monney, C. and De Giovannini, U. and Puppin, M. and Hoesch, M. and Springate, E. and Chapman, R. T. and Cacho, C. and Wolf, M. and Rubio, A. and Ernstorfer, R.},
  journal = {Phys. Rev. Lett.},
  volume = {117},
  issue = {27},
  pages = {277201},
  numpages = {5},
  year = {2016},
  month = {Dec},
  publisher = {American Physical Society},
  doi = {10.1103/PhysRevLett.117.277201},
  url = {https://link.aps.org/doi/10.1103/PhysRevLett.117.277201}
}

@article{Madeo2020,
author = {Julien Madéo  and Michael K. L. Man  and Chakradhar Sahoo  and Marshall Campbell  and Vivek Pareek  and E. Laine Wong  and Abdullah Al-Mahboob  and Nicholas S. Chan  and Arka Karmakar  and Bala Murali Krishna Mariserla  and Xiaoqin Li  and Tony F. Heinz  and Ting Cao  and Keshav M. Dani },
title = {Directly visualizing the momentum-forbidden dark excitons and their dynamics in atomically thin semiconductors},
journal = {Science},
volume = {370},
number = {6521},
pages = {1199-1204},
year = {2020},
doi = {10.1126/science.aba1029},
}

@article{Fanciulli2023,
  title = {Ultrafast Hidden Spin Polarization Dynamics of Bright and Dark Excitons in {2H}{-}{{WSe}}{\textsubscript{2}}},
  author = {Fanciulli, Mauro and Bresteau, David and Gaudin, J\'er\^ome and Dong, Shuo and G\'eneaux, Romain and Ruchon, Thierry and Tcherbakoff, Olivier and Min\'ar, J\'an and Heckmann, Olivier and Richter, Maria Christine and Hricovini, Karol and Beaulieu, Samuel},
  journal = {Phys. Rev. Lett.},
  volume = {131},
  issue = {6},
  pages = {066402},
  numpages = {6},
  year = {2023},
  month = {Aug},
  publisher = {American Physical Society},
  doi = {10.1103/PhysRevLett.131.066402},
  url = {https://link.aps.org/doi/10.1103/PhysRevLett.131.066402},
}

@Article{Trovatello2020,
author={Trovatello, Chiara
and Katsch, Florian
and Borys, Nicholas J.
and Selig, Malte
and Yao, Kaiyuan
and Borrego-Varillas, Rocio
and Scotognella, Francesco
and Kriegel, Ilka
and Yan, Aiming
and Zettl, Alex
and Schuck, P. James
and Knorr, Andreas
and Cerullo, Giulio
and Conte, Stefano Dal},
title={{The ultrafast onset of exciton formation in 2D semiconductors}},
journal={Nature Communications},
year={2020},
month={Oct},
day={19},
volume={11},
number={1},
pages={5277},
issn={2041-1723},
doi={10.1038/s41467-020-18835-5},
url={https://doi.org/10.1038/s41467-020-18835-5}
}

@ARTICLE{Gosetti2025,
   author = {Gosetti, V. and Cervantes-Villanueva, J. and Mor, S. and Sangalli, D. and Garcia-Cristobal, A. and Molina-Sanchez, A. and Agekyan, V.F. and Tuniz, M. and Puntel, D. and Bronsch, W. and Cilento, F. and Pagliara, S.},
   title = {Unveiling the exciton formation in time, energy and momentum domain in layered van der {Waals} semiconductors},
   journal = {Progress in Surface Science},
   volume = {100},
   pages = {100777},
   year = {2025},
   doi={https://doi.org/10.1016/j.progsurf.2025.100777}
}

@ARTICLE{Carre2021,
   author = {Carr\'e, E. and Sponza, L. and Lusson, A. and Stenger, I. and Gaufr\`es, E. and Loiseau, A. and Barjon, J.},
   title = {Excitons in bulk black phosphorus evidenced by photoluminescence at low temperature},
   journal = {2D Materials},
   volume = {8},
   pages = {021001},
   year = {2021},
   doi={https://doi.org/10.1088/2053-1583/abca81}
}

@ARTICLE{Dong2021,
   author = {Dong, S. and Puppin, M. and Pincelli, T. and Beaulieu, S. and
Christiansen, C. and Huebener, H. and Nicholso, C.W. and Xian, R.P. and Dendzik, M. and Deng, Y. and Windsor, Y.W. and Selig, M. and Malic, E. and Rubio, A. and Knorr, A. and Wolf, M. and Rettig, L. and Ernstorfer, R.},
   title = {{Directmeasurement of key exciton properties: Energy,
dynamics, and spatial distribution of the wave function}},
   journal = {Natural Sciences},
   volume = {1},
   pages = {e10010.},
   year = {2021},
   doi={https://doi.org/10.1002/ntls.10010}
}

@article{rustagi_photoemission_2018,
   title       = {Photoemission signature of excitons},
   volume      = {97},
   url         = {https://link.aps.org/doi/10.1103/PhysRevB.97.235310},
   doi         = {10.1103/PhysRevB.97.235310},
   number      = {23},
   journal     = {Phys. Rev. B},
   publisher   = {American Physical Society},
   author      = {Rustagi, Avinash and Kemper, Alexander F.},
   year        = {2018},
   pages       = {235310},
}

@article{faure_full_2012,
	title = {Full characterization and optimization of a femtosecond ultraviolet laser source for time and angle-resolved photoemission on solid surfaces},
	volume = {83},
	issn = {0034-6748},
	url = {https://doi.org/10.1063/1.3700190},
	doi = {10.1063/1.3700190},
	pages = {043109},
	number = {4},
	journal = {Review of Scientific Instruments},
	shortjournal = {Review of Scientific Instruments},
	author = {Faure, J. and Mauchain, J. and Papalazarou, E. and Yan, W. and Pinon, J. and Marsi, M. and Perfetti, L.},
	urldate = {2025-04-08},
	date = {2012-04-13},
        year = {2012},
}

@article{chen_band_2019,
	title = {Band {Gap} {Renormalization}, {Carrier} {Multiplication}, and {Stark} {Broadening} in {Photoexcited} {Black} {Phosphorus}},
	volume = {19},
	issn = {1530-6984, 1530-6992},
	url = {https://pubs.acs.org/doi/10.1021/acs.nanolett.8b04344},
	doi = {10.1021/acs.nanolett.8b04344},
	number = {1},
	urldate = {2025-07-10},
	journal = {Nano Letters},
	author = {Chen, Zhesheng and Dong, Jingwei and Papalazarou, Evangelos and Marsi, Marino and Giorgetti, Christine and Zhang, Zailan and Tian, Bingbing and Rueff, Jean-Pascal and Taleb-Ibrahimi, Amina and Perfetti, Luca},
	month = jan,
	year = {2019},
	pages = {488--493},
}

@article{shen_ultrafast_2025,
	title = {Ultrafast energizing the parity-forbidden dark exciton in black phosphorus},
	volume = {16},
	copyright = {2025 The Author(s)},
	issn = {2041-1723},
	url = {https://www.nature.com/articles/s41467-025-58930-z},
	doi = {10.1038/s41467-025-58930-z},
	number = {1},
	urldate = {2026-02-26},
	journal = {Nature Communications},
	publisher = {Nature Publishing Group},
	author = {Shen, Guangzhen and Tian, Xirui and Cao, Limin and Guo, Hongli and Li, Xintong and Tian, Yishu and Cui, Xuefeng and Feng, Min and Zhao, Jin and Wang, Bing and Petek, Hrvoje and Tan, Shijing},
	month = apr,
	year = {2025},
	pages = {3992},
}

@article{kremer_ultrafast_2021,
	title = {Ultrafast dynamics of the surface photovoltage in potassium-doped black phosphorus},
	volume = {104},
	url = {https://link.aps.org/doi/10.1103/PhysRevB.104.035125},
	doi = {10.1103/PhysRevB.104.035125},
	number = {3},
	urldate = {2026-02-26},
	journal = {Physical Review B},
	publisher = {American Physical Society},
	author = {Kremer, G. and Rumo, M. and Yue, C. and Pulkkinen, A. and Nicholson, C. W. and Jaouen, T. and von Rohr, F. O. and Werner, P. and Monney, C.},
	month = jul,
	year = {2021},
	pages = {035125},
}

@article{hedayat_non-equilibrium_2021,
	title = {Non-equilibrium band broadening, gap renormalization and band inversion in black phosphorus},
	volume = {8},
	issn = {2053-1583},
	url = {https://iopscience.iop.org/article/10.1088/2053-1583/abd89a},
	doi = {10.1088/2053-1583/abd89a},
	number = {2},
	urldate = {2025-07-10},
	journal = {2D Materials},
	author = {Hedayat, H and Ceraso, A and Soavi, G and Akhavan, S and Cadore, A and Dallera, C and Cerullo, G and Ferrari, A C and Carpene, E},
	month = apr,
	year = {2021},
	pages = {025020},
}

@article{roth_photocarrier-induced_2019,
	title = {Photocarrier-induced band-gap renormalization and ultrafast charge dynamics in black phosphorus},
	volume = {6},
	issn = {2053-1583},
	url = {https://doi.org/10.1088/2053-1583/ab1216},
	doi = {10.1088/2053-1583/ab1216},
	number = {3},
	urldate = {2026-02-27},
	journal = {2D Materials},
	publisher = {IOP Publishing},
	author = {Roth, S and Crepaldi, A and Puppin, M and Gatti, G and Bugini, D and Grimaldi, I and Barrilot, T R and Arrell, C A and Frassetto, F and Poletto, L and Chergui, M and Marini, A and Grioni, M},
	month = apr,
	year = {2019},
	pages = {031001},
}

@article{golias2016disentangling,
  title={Disentangling bulk from surface contributions in the electronic structure of black phosphorus},
  author={Golias, E and Krivenkov, M and S{\'a}nchez-Barriga, J},
	url = {https://doi.org/10.1103/PhysRevB.93.075207},
  journal={Phys. Rev. B},
  volume={93},
  number={7},
  pages={075207},
  year={2016},
  publisher={APS}
}

@article{man_experimental_2021,
author = {Michael K. L. Man  and Julien Madéo  and Chakradhar Sahoo  and Kaichen Xie  and Marshall Campbell  and Vivek Pareek  and Arka Karmakar  and E Laine Wong  and Abdullah Al-Mahboob  and Nicholas S. Chan  and David R. Bacon  and Xing Zhu  and Mohamed M. M. Abdelrasoul  and Xiaoqin Li  and Tony F. Heinz  and Felipe H. da Jornada  and Ting Cao  and Keshav M. Dani },
title = {Experimental measurement of the intrinsic excitonic wave function},
journal = {Science Advances},
volume = {7},
number = {17},
pages = {eabg0192},
year = {2021},
doi = {10.1126/sciadv.abg0192},
}

@article{onida_electronic_2002,
   title       = {Electronic excitations: density-functional versus many-body {Green}’s-function approaches},
   volume      = {74},
   url         = {http://link.aps.org/doi/10.1103/RevModPhys.74.601},
   doi         = {10.1103/RevModPhys.74.601},
   number      = {2},
   journal     = {Rev. Mod. Phys.},
   author      = {Onida, Giovanni and Reining, Lucia and Rubio, Angel},
   year        = {2002},
   pages       = {601--659},
}

@article{sangalli_ab-initio_2018,
   title       = {An ab-initio approach to describe coherent and non-coherent exciton dynamics},
   volume      = {91},
   url         = {https://link.springer.com/article/10.1140/epjb/e2018-90126-5},
   doi         = {10.1140/epjb/e2018-90126-5},
   number      = {8},
   journal     = {Eur. Phys. J. B},
   author      = {Sangalli, Davide and Perfetto, Enrico and Stefanucci, Gianluca and Marini, Andrea},
   year        = {2018},
   pages       = {171},
}

@article{boschini_time-resolved_2024,
   title       = {Time-resolved {ARPES} studies of quantum materials},
   volume      = {96},
   url         = {https://link.aps.org/doi/10.1103/RevModPhys.96.015003},
   doi         = {10.1103/RevModPhys.96.015003},
   number      = {1},
   journal     = {Rev. Mod. Phys.},
   author      = {Boschini, Fabio and Zonno, Marta and Damascelli, Andrea},
   year        = {2024},
   pages       = {015003},
}

@article{pizzi_wannier90_2020,
   title       = {Wannier90 as a community code: new features and applications},
   volume      = {32},
   url         = {https://iopscience.iop.org/article/10.1088/1361-648X/ab51ff},
   doi         = {10.1088/1361-648X/ab51ff},
   number      = {16},
   journal     = {J. Phys.: Condens. Matter},
   author      = {Pizzi, Giovanni and Vitale, Valerio and Arita, Ryotaro and Bl\"{u}gel, Stefan and Freimuth, Frank and G\'{e}ranton, Guillaume and Gibertini, Marco and Gresch, Dominik and Johnson, Charles and Koretsune, Takashi and Iba\~{n}ez-Azpiroz, Julen and Lee, Hyungjun and Lihm, Jae-Mo and Marchand, Daniel and Marrazzo, Antimo and Mokrousov, Yuriy and Mustafa, Jamal I and Nohara, Yoshiro and Nomura, Yusuke and Paulatto, Lorenzo and Ponc\'{e}, Samuel and Ponweiser, Thomas and Qiao, Junfeng and Th\"{o}le, Florian and Tsirkin, Stepan S and Wierzbowska, Małgorzata and Marzari, Nicola and Vanderbilt, David and Souza, Ivo and Mostofi, Arash A and Yates, Jonathan R},
   year        = {2020},
   pages       = {165902},
}

@article{wang_colloquium_2018,
   title       = {\textit{{Colloquium}} : {Excitons} in atomically thin transition metal dichalcogenides},
   volume      = {90},
   url         = {https://link.aps.org/doi/10.1103/RevModPhys.90.021001},
   doi         = {10.1103/RevModPhys.90.021001},
   number      = {2},
   journal     = {Rev. Mod. Phys.},
   author      = {Wang, Gang and Chernikov, Alexey and Glazov, Mikhail M. and Heinz, Tony F. and Marie, Xavier and Amand, Thierry and Urbaszek, Bernhard},
   year        = {2018},
   pages       = {021001},
}

@article{kira_many-body_2006,
   title       = {Many-body correlations and excitonic effects in semiconductor spectroscopy},
   volume      = {30},
   copyright   = {https://www.elsevier.com/tdm/userlicense/1.0/},
   url         = {https://linkinghub.elsevier.com/retrieve/pii/S0079672706000280},
   doi         = {10.1016/j.pquantelec.2006.12.002},
   number      = {5},
   journal     = {Prog. Quantum Electron.},
   author      = {Kira, M. and Koch, S.W.},
   year        = {2006},
   pages       = {155--296},
}

@article{lindberg_effective_1988,
   title       = {Effective {Bloch} equations for semiconductors},
   volume      = {38},
   copyright   = {http://link.aps.org/licenses/aps-default-license},
   url         = {https://link.aps.org/doi/10.1103/PhysRevB.38.3342},
   doi         = {10.1103/PhysRevB.38.3342},
   number      = {5},
   journal     = {Phys. Rev. B},
   author      = {Lindberg, M. and Koch, S. W.},
   year        = {1988},
   pages       = {3342--3350},
}

@article{dias_wantibexos_2023,
   title       = {{WanTiBEXOS}: {A} {Wannier} based {Tight} {Binding} code for electronic band structure, excitonic and optoelectronic properties of solids},
   volume      = {285},
   url         = {https://linkinghub.elsevier.com/retrieve/pii/S0010465522003551},
   doi         = {10.1016/j.cpc.2022.108636},
   journal     = {Comp. Phys. Commun.},
   author      = {Dias, Alexandre C. and Silveira, Julian F.R.V. and Qu, Fanyao},
   year        = {2023},
   pages       = {108636},
}

@article{antonius_theory_2022,
   title       = {Theory of exciton-phonon coupling},
   volume      = {105},
   url         = {https://link.aps.org/doi/10.1103/PhysRevB.105.085111},
   doi         = {10.1103/PhysRevB.105.085111},
   number      = {8},
   journal     = {Phys. Rev. B},
   author      = {Antonius, Gabriel and Louie, Steven G.},
   year        = {2022},
   pages       = {085111},
}

@article{rohlfing_electron-hole_2000,
   title       = {Electron-hole excitations and optical spectra from first principles},
   volume      = {62},
   copyright   = {http://link.aps.org/licenses/aps-default-license},
   url         = {https://link.aps.org/doi/10.1103/PhysRevB.62.4927},
   doi         = {10.1103/PhysRevB.62.4927},
   number      = {8},
   journal     = {Phys. Rev. B},
   author      = {Rohlfing, Michael and Louie, Steven G.},
   year        = {2000},
   pages       = {4927--4944},
}

@article{zahn_anisotropic_2020,
   title       = {Anisotropic {Nonequilibrium} {Lattice} {Dynamics} of {Black} {Phosphorus}},
   volume      = {20},
   copyright   = {http://pubs.acs.org/page/policy/authorchoice\_ccby\_termsofuse.html},
   url         = {https://pubs.acs.org/doi/10.1021/acs.nanolett.0c00734},
   doi         = {10.1021/acs.nanolett.0c00734},
   number      = {5},
   journal     = {Nano Lett.},
   author      = {Zahn, Daniela and Hildebrandt, Patrick-Nigel and Vasileiadis, Thomas and Windsor, Yoav William and Qi, Yingpeng and Seiler, H\'{e}l\`{e}ne and Ernstorfer, Ralph},
   year        = {2020},
   pages       = {3728--3733},
}

@article{zhang_determination_2018,
   title       = {Determination of layer-dependent exciton binding energies in few-layer black phosphorus},
   volume      = {4},
   url         = {https://www.science.org/doi/10.1126/sciadv.aap9977},
   doi         = {10.1126/sciadv.aap9977},
   number      = {3},
   journal     = {Sci. Adv.},
   author      = {Zhang, Guowei and Chaves, Andrey and Huang, Shenyang and Wang, Fanjie and Xing, Qiaoxia and Low, Tony and Yan, Hugen},
   year        = {2018},
   pages       = {eaap9977},
}

@article{tran_layer-controlled_2014,
   title       = {Layer-controlled band gap and anisotropic excitons in few-layer black phosphorus},
   volume      = {89},
   copyright   = {http://link.aps.org/licenses/aps-default-license},
   url         = {https://link.aps.org/doi/10.1103/PhysRevB.89.235319},
   doi         = {10.1103/PhysRevB.89.235319},
   number      = {23},
   journal     = {Phys. Rev. B},
   author      = {Tran, Vy and Soklaski, Ryan and Liang, Yufeng and Yang, Li},
   year        = {2014},
   pages       = {235319},
}

@article{chan_excitonphonon_2024,
   title       = {Exciton-{Phonon} {Coupling} {Induces} a {New} {Pathway} for {Ultrafast} {Intralayer}-to-{Interlayer} {Exciton} {Transition} and {Interlayer} {Charge} {Transfer} in {WS}$_{\textrm{2}}$ -{MoS}$_{\textrm{2}}$ {Heterostructure}: {A} {First}-{Principles} {Study}},
   volume      = {24},
   copyright   = {https://creativecommons.org/licenses/by/4.0/},
   url         = {https://pubs.acs.org/doi/10.1021/acs.nanolett.4c01508},
   doi         = {10.1021/acs.nanolett.4c01508},
   number      = {26},
   journal     = {Nano Lett.},
   author      = {Chan, Yang-hao and Naik, Mit H. and Haber, Jonah B. and Neaton, Jeffrey B. and Louie, Steven G. and Qiu, Diana Y. and Da Jornada, Felipe H.},
   year        = {2024},
   pages       = {7972--7978},
}

@article{reutzel_probing_2024,
   title       = {Probing excitons with time-resolved momentum microscopy},
   volume      = {9},
   url         = {https://www.tandfonline.com/doi/full/10.1080/23746149.2024.2378722},
   doi         = {10.1080/23746149.2024.2378722},
   number      = {1},
   journal     = {Advances in Physics: X},
   author      = {Reutzel, Marcel and Jansen, G. S. Matthijs and Mathias, Stefan},
   year        = {2024},
   pages       = {2378722},
}

@article{wang_highly_2015,
   title       = {Highly anisotropic and robust excitons in monolayer black phosphorus},
   volume      = {10},
   url         = {https://www.nature.com/articles/nnano.2015.71},
   doi         = {10.1038/nnano.2015.71},
   number      = {6},
   journal     = {Nature Nanotech},
   author      = {Wang, Xiaomu and Jones, Aaron M. and Seyler, Kyle L. and Tran, Vy and Jia, Yichen and Zhao, Huan and Wang, Han and Yang, Li and Xu, Xiaodong and Xia, Fengnian},
   year        = {2015},
   pages       = {517--521},
}

@article{montanaro_anomalous_2022,
   title       = {Anomalous non-equilibrium response in black phosphorus to sub-gap mid-infrared excitation},
   volume      = {13},
   url         = {https://www.nature.com/articles/s41467-022-30341-4},
   doi         = {10.1038/s41467-022-30341-4},
   number      = {1},
   journal     = {Nat Commun},
   author      = {Montanaro, Angela and Giusti, Francesca and Zanfrognini, Matteo and Di Pietro, Paola and Glerean, Filippo and Jarc, Giacomo and Rigoni, Enrico Maria and Mathengattil, Shahla Y. and Varsano, Daniele and Rontani, Massimo and Perucchi, Andrea and Molinari, Elisa and Fausti, Daniele},
   year        = {2022},
   pages       = {2667},
}

@article{chan_exciton_2023,
   title       = {Exciton {Lifetime} and {Optical} {Line} {Width} {Profile} via {Exciton}-{Phonon} {Interactions}: {Theory} and {First}-{Principles} {Calculations} for {Monolayer} {MoS}$_{\textrm{2}}$},
   volume      = {23},
   copyright   = {https://doi.org/10.15223/policy-029},
   url         = {https://pubs.acs.org/doi/10.1021/acs.nanolett.3c00732},
   doi         = {10.1021/acs.nanolett.3c00732},
   number      = {9},
   journal     = {Nano Lett.},
   author      = {Chan, Yang-hao and Haber, Jonah B. and Naik, Mit H. and Neaton, Jeffrey B. and Qiu, Diana Y. and Da Jornada, Felipe H. and Louie, Steven G.},
   year        = {2023},
   pages       = {3971--3977},
}

@article{bieniek_theory_2022,
   title       = {Theory of {Excitons} in {Atomically} {Thin} {Semiconductors}: {Tight}-{Binding} {Approach}},
   volume      = {12},
   url         = {https://www.mdpi.com/2079-4991/12/9/1582},
   doi         = {10.3390/nano12091582},
   number      = {9},
   journal     = {Nanomaterials},
   author      = {Bieniek, Maciej and Sadecka, Katarzyna and Szulakowska, Ludmiła and Hawrylak, Paweł},
   year        = {2022},
   pages       = {1582},
}

@book{Klingshirn2012,
  author    = {Klingshirn, Claus F.},
  title     = {Semiconductor Optics},
  year      = {2012},
  publisher = {Springer Berlin Heidelberg},
  address   = {Berlin, Heidelberg},
  edition   = {4th},
  pages     = {491--597},
  note      = {Chaps. 19--21},
  isbn      = {978-3-642-28362-8},
  doi       = {10.1007/978-3-642-28362-8}
}

@article{aryasetiawan_frequency-dependent_2004,
   title       = {Frequency-dependent local interactions and low-energy effective models from electronic structure calculations},
   volume      = {70},
   url         = {https://link.aps.org/doi/10.1103/PhysRevB.70.195104},
   doi         = {10.1103/PhysRevB.70.195104},
   number      = {19},
   journal     = {Phys. Rev. B},
   publisher   = {American Physical Society},
   author      = {Aryasetiawan, F. and Imada, M. and Georges, A. and Kotliar, G. and Biermann, S. and Lichtenstein, A. I.},
   year        = {2004},
   pages       = {195104},
}

@misc{dendzik_ultrafast_2026,
   title       = {Ultrafast nonlinear {Hall} effect in black phosphorus},
   copyright   = {Creative Commons Attribution 4.0 International},
   url         = {https://arxiv.org/abs/2604.06083},
   urldate     = {2026-04-10},
   publisher   = {arXiv},
   author      = {Dendzik, Maciej and Marini, Andrea and Beaulieu, Samuel and Dong, Shuo and Pincelli, Tommaso and Maklar, Julian and Xian, R. Patrick and Perfetto, Enrico and Wolf, Martin and Stefanucci, Gianluca and Ernstorfer, Ralph and Rettig, Laurenz},
   year        = {2026},
}

@Article{Stefanucci_ExcitonicBlochEquations_2025,
	title={{Excitonic Bloch equations from first principles}},
	author={Gianluca Stefanucci and Enrico Perfetto},
	journal={SciPost Phys.},
	volume={18},
	pages={009},
	year={2025},
	publisher={SciPost},
	doi={10.21468/SciPostPhys.18.1.009},
	url={https://scipost.org/10.21468/SciPostPhys.18.1.009},
}

\section*{Acknowledgments}

This project was supported by Swiss National Science Foundation Grant No. 10000782 and No. P00P2\_170597. Skillful technical assistance was provided by J. L. Andrey, M. Andrey, F. Bourqui, and B. Hediger.

\section*{Author contributions statement}

C.M., M.S. and G.K. designed the research. G.K., J.M., A.M., F.C. and M.R. carried out the experiments and measurements. V.C. and P.W. performed the DFT calculations. F.O.v.R. and M. B. performed sample growth and initial characterization. G.K., J.M. and A.M. analyzed the data with help from all the authors. J.F.P.M. and M.S. developped the quantum-kinetic theoretical model and performed the simulations. J.F.P.M., J.M., C.M., M.S. and G.K. wrote the paper, with input from all authors. All authors reviewed the manuscript.

\newpage
\onecolumngrid

\clearpage
\setcounter{page}{1}
\renewcommand{\thepage}{S\arabic{page}}
\setcounter{figure}{0}
\renewcommand{\thefigure}{S\arabic{figure}}
\renewcommand{\theHfigure}{S\arabic{figure}} 

\setcounter{equation}{0}
\renewcommand{\theequation}{S\arabic{equation}}
\renewcommand{\theHequation}{S\arabic{equation}}

{\Large {\begin{center}
\textit{Supplementary Material for ``Ultrafast dynamics of excitons in black phosphorus''}\end{center}}}

\vspace{1cm}

\section{Band Structure and Wannier-Bloch basis}
\begin{figure}[h]
    \centering\includegraphics[scale=0.7]{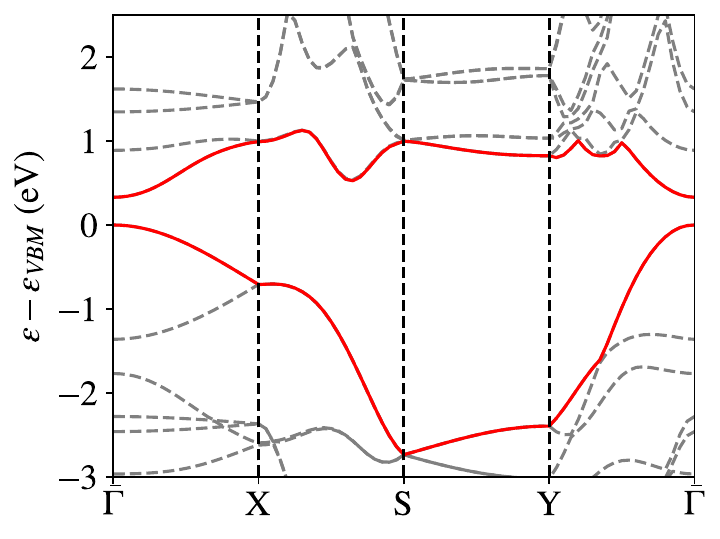}
    \caption{ \textbf{Band Structure of Bulk Black Phosphorus}:
    $G^0W^0$ quasi-particle band structure (gray dashed lines). The two red lines show the top valence and bottom conduction band, respectively.
    }
    \label{fig:band_structure}
\end{figure}

We performed density functional theory (DFT) calculations using the FLEUR full-potential all-electron code \cite{wortmann_2023_7778444}.
We used a $12 \times 12 \times 8$ $\mathbf{k}$-grid for the calculations.
The atomic positions and lattice parameters for the orthorhombic lattice are given in Table~\ref{tab:lattice}.
\begin{table}[h!]
    \begin{tabular}{c|c|c}
        $a$      &     $b$      &     $c$      \\ \hline
        $3.331$ \AA & $ 4.585$ \AA & $10.989$ \AA  \\ \hline
    \end{tabular}
    \caption{Lattice parameters (in \AA) for the relaxed structure of bulk black Phosphorus.}
    \label{tab:lattice}
\end{table}
On top of the DFT calculations, we performed a one-shot $G^0W^0$ calculation \cite{Hedin1965} to obtain the self-energy $\Sigma_{\alpha}(\mathbf{k})$, from which the quasi-particle energies $\varepsilon^{QP}_{\alpha}$ are computed.
The quasi-particle energies are then obtained by solving the quasi-particle equation
\begin{equation}
    \varepsilon_{\alpha}^{QP} (\mathbf{k}) = 
    \varepsilon_{\alpha}(\mathbf{k}) - V_{\alpha}^{xc} (\mathbf{k}) + 
    \Re\left\{ \Sigma_{\alpha} (\mathbf{k},\varepsilon_{\alpha}^{QP} (\mathbf{k})) \right\},
\end{equation}
where $\varepsilon_{\alpha}(\mathbf{k})$ and $V_{\alpha}^{xc}(\mathbf{k})$ are the Kohn-Sham eigenvalues and the exchange-correlation potential from the DFT calculation, respectively.
The resulting quasi-particle Hamiltonian is expressed in the basis of projective Wannier functions $\phi_{j} (\mathbf{r} - \mathbf{R})$ using the Wannier90 code \cite{Pizzi2020}. We include  the $s$- and $p$-orbitals of Phosphorus.
The band structure of the system is shown in Fig.~\ref{fig:band_structure}.
We shifted the bands such that the valence band maximum (VBM) is located at $\varepsilon_\mathrm{VBM}=0$.
Because the $G_0 W_0$ bands predict a too large band gap compared to the experimental value, we applied a scissor operator to fix, $E_{gap} = 0.31$ eV. In the equations and discussion listed below, we have omitted the labels $QP$ in the quasi-particle states.

\section{Coulomb interactions}

As the next step, we performed constrained random-phase approximation (cRPA) calculations to obtain the Coulomb matrix elements in the Wannier basis using the SPEX code \cite{Friedrich2010, Aryasetiawan2004}, with the target space being the valence and conduction bands.
Keeping only the density-density matrix elements due to the localized nature of the Wannier functions and the static approximation, the interaction Hamiltonian takes the form
\begin{equation}
    \hat{H}_{int} = 
    \frac{1}{2} \sum_{\mathbf{R} \mathbf{R}'} \sum_{ij} U_{ij}(\mathbf{R}- \mathbf{R}') \hat{n}_{i\mathbf{R}} \hat{n}_{j\mathbf{R}'},
    \label{eq:interaction_hamiltonian}
\end{equation}
where $\hat{n}_{i\mathbf{R}} = \hat{c}^{\dagger}_{i \mathbf{R}} \hat{c}_{i \mathbf{R}}$ is the density operator for orbital $i$ on lattice site $\mathbf{R}$.

The cRPA calculations have been performed on a $N=32\times32\times32$ supercell, which is sufficient to accurately describe the short-range part of $U_{ij}(\mathbf{R})$. To capture the asymptotic behavior of the effective bare Coulomb
interaction, we have fitted it to a function of the shape
\begin{equation}
    \label{eq:bare_coulomb}
    U_{ij}(\mathbf{R}) \propto \frac{v_{ij}}{\left| \mathbf{R} + (\mathbf{r}_i - \mathbf{r}_j) \right|} \ .
\end{equation}
From the representation in the Wannier basis of the effective bare Coulomb interaction, we compute Coulomb interaction in the band basis by
\begin{equation}
    V_{\alpha\beta\gamma\sigma}(\mathbf{k}, \mathbf{k}^\prime, \mathbf{Q}) = \sum_{ij} U_{ij}(\mathbf{Q}) 
    C_{i\alpha}^{*}(\mathbf{k}) C_{j\beta}^{*}(\mathbf{k}') C_{j\alpha}^{*}(\mathbf{k}' + \mathbf{Q}) C_{i\beta}(\mathbf{k} - \mathbf{Q}) \ .
\end{equation}

For computing the screened interaction, we performed RPA calculations. However, since the convergence is slow and computationally expensive, we used the information from the RPA calculation for the short-range part, but used a model dielectric function $\epsilon(\mathbf{Q})$ (we only consider static screening) for the long-range interaction. The long-range part screened interaction in Wannier basis is then defined by
\begin{equation}
    W_{ij}(\mathbf{Q}) = \epsilon^{-1}(\mathbf{Q},\omega=0) U_{ij}(\mathbf{Q}) \ .
    \label{eq:screened}
\end{equation}
For bulk semiconductors, a well-used model was proposed in Ref. \cite{Cappellini1993} were the static dielectric function takes the form
\begin{equation}
    \epsilon (\mathbf{Q}) = 1 + \frac{1}
    { \frac{1}{\epsilon(Q=0)-1} + \alpha \frac{Q^2}{q_{TF}^2} + \frac{Q^4}{4\omega_p} \ .
    },
\end{equation}
Here, $q_{TF} = 2(3\bar{n}/\pi )^{1/6}$ and $\omega_p = \sqrt{4\pi\bar{n}}$ (where $\bar{n}=n/V$ is the average density of electrons) are the Thomas-Fermi wave vector and the plasma frequency, respectively. 
For this case, the number of electrons in the system is $48$, and the volume $V$ is given by the product of the lattice vectors in Table~\ref{tab:lattice}.
The parameter $\alpha$ is set to $1.563$, which is the averaged value extracted from other materials. Small variations of it do not drastically affect the RPA results \cite{Cappellini1993}.
The value we used for the macroscopic dielectric constant is $\epsilon(\mathbf{Q} = 0) = 6.36$, obtained by minimizing the error of the screened interaction with the RPA calculation in the short range.
By using the models described in Eq.~\eqref{eq:bare_coulomb} and Eq.~\eqref{eq:screened}, we ensure the asymptotic behavior of the interactions.

\begin{figure}[h]
    \centering\includegraphics[width=0.8\textwidth]{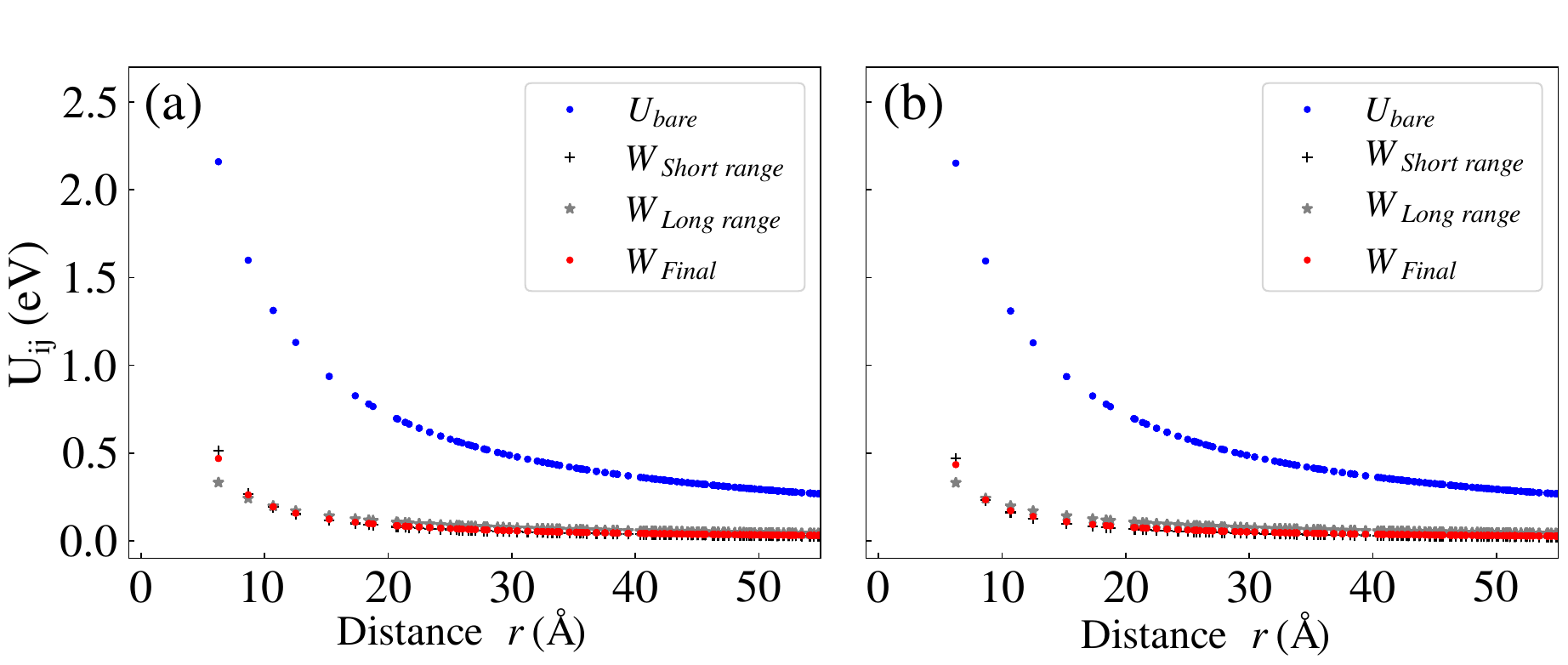}
    \caption{ \textbf{Distance Dependence of the Effective Interactions}:
    In blue, the static nonlocal cRPA interaction $U _{ij}(\mathbf{R})$ fitted to the model function given in Eq. \eqref{eq:bare_coulomb}. In black and gray, the short- and long-range screened Coulomb interactions. The red dots are the screened interactions obtained by merging the short- and long-range components. Panels (a) and (b) show the interband interactions between Wannier orbitals with indexes $i,j = 19 $ and $i,j = 20$, respectively, dominated by the $p_z$ orbitals.
    }
    \label{fig:interactions}
\end{figure}

In Fig.~\ref{fig:interactions}, the short- and long-range screened interactions are shown with black and gray dots, respectively.
The red dots correspond to the screened interaction used in the following sections, which is obtained by merging the short- and long-range ones.
For distances smaller than their intersection, we used the short-range fitting, and for distances larger than the intersection point, we used the long-range interaction.

\section{Bethe-Salpeter equation (BSE)}

\begin{figure}[h]
    \centering\includegraphics[scale=0.5]{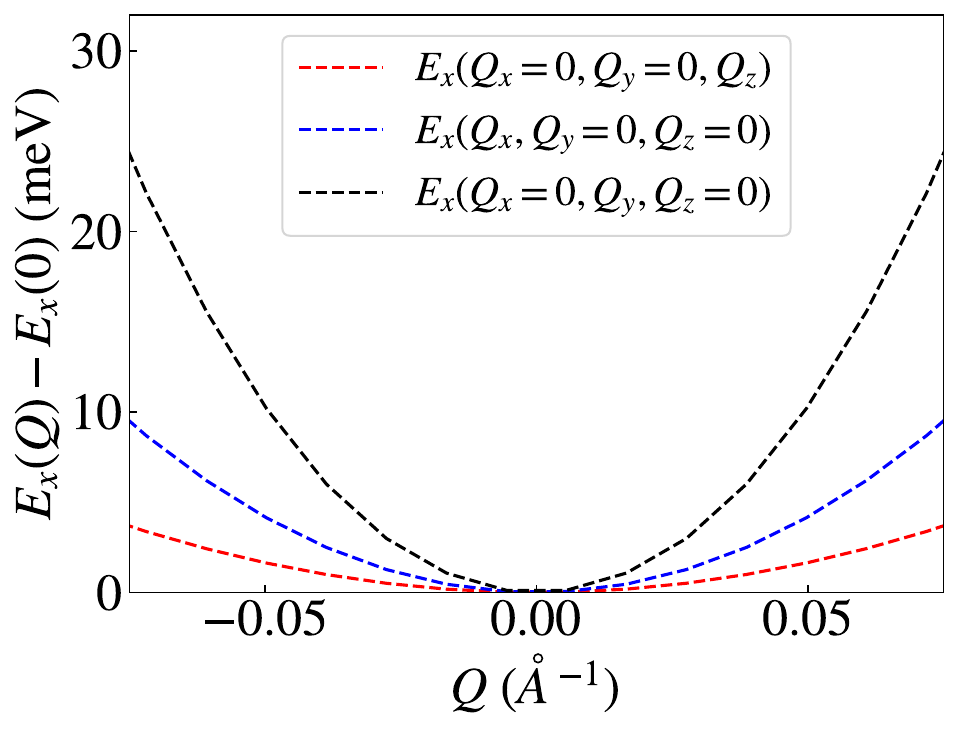}
    \caption{ \textbf{Exciton Dispersion Energy}: Ground state dispersion energy $E_x$ along the different directions $Q_x$, $Q_y$, and $Q_z$.
    }
    \label{fig:exitons_dispersion}
\end{figure}

After computing the band structure, the bare Coulomb, and the screened interaction, we proceed to computing the excitons properties.
The exciton state can be written as
\begin{equation}
    \left| \Psi^{\lambda,\mathbf{Q}} \right> = 
    \frac{1}{N_{k}}\sum_{\mathbf{k}} \sum_{\zeta\nu} \Lambda_{\nu\zeta}^{\lambda,\mathbf{Q}} (\mathbf{k}) 
    \hat{c}_{\nu,\mathbf{k} + \mathbf{Q} }^{\dagger} \hat{v}_{\zeta,\mathbf{k}} \left| \Psi_0 \right>,
\end{equation}
where $\Lambda_{\nu\zeta}^{\lambda,\mathbf{Q}}(\mathbf{k})$ is the exciton wave function, whose quantum numbers we label by $\lambda$.
The operators $\hat{v}_{\zeta,\mathbf{k}}$ remove an electron in the valence ($\zeta$) states, and the operators $\hat{c}_{\nu,\mathbf{k} + \mathbf{Q} }^{\dagger}$ create an electron in the conduction ($\nu$) states with momentum $\mathbf{k}+\mathbf{Q}$.
To obtain the exciton wave function, we solved the BSE, which reads as
\begin{equation}
    \left( 
        E_{\lambda} \left( \mathbf{Q} \right) -\varepsilon_C%(\mathbf{k} + \mathbf{Q}) 
        + \varepsilon_V %(\mathbf{k})
    \right) \Lambda_{CV}^{\lambda,\mathbf{Q}} 
    \left(
        \mathbf{k}
    \right) =
    - \sum_{C' V'} \left[ 
        W_{C V' V C'} - V_{C V' C' V}
    \right] \Lambda_{C' V'}^{\lambda,\mathbf{Q}} \left(
        \mathbf{k}'
    \right),
    \label{eq:BSE}
\end{equation}
where the index $C$ ($C'$) corresponds to the conduction state with momentum $\mathbf{k} + \mathbf{Q}$ ($\mathbf{k}' + \mathbf{Q}$). The index $V$ ($V'$)
corresponds to the valence state with momentum $\mathbf{k}$ ($\mathbf{k}'$). In our case, we are just taking into account one single valence and conduction state.
% , leaving just the sum over momentum $\mathbf{k}'$ on the right side of the equality.
From now on we omit the labels $C$ and $V$ in the exciton wave function $\left[\Lambda^{\lambda,\mathbf{Q}} \left( \mathbf{k}\right)\right]$.
The corresponding dispersion energy for the ground-state excitons ($E_{\lambda=0}$) is shown in Fig.~\ref{fig:exitons_dispersion} along the $Q_x$, $Q_y$, and $Q_z$ directions. 
Note that the band gap of the band dispersion energy does not change the exciton dispersion energy, only the binding energy.
Therefore, we plotted it by setting the zero energy to the coherent exciton energy.% $E_x(\mathbf{Q}=0) = 0.29$ eV.
In this study we have focused on the ground state exciton ($\lambda=0$), thus in the analysis we have skipped these indexes.

\begin{figure}[h]
    \centering\includegraphics[scale=0.4]{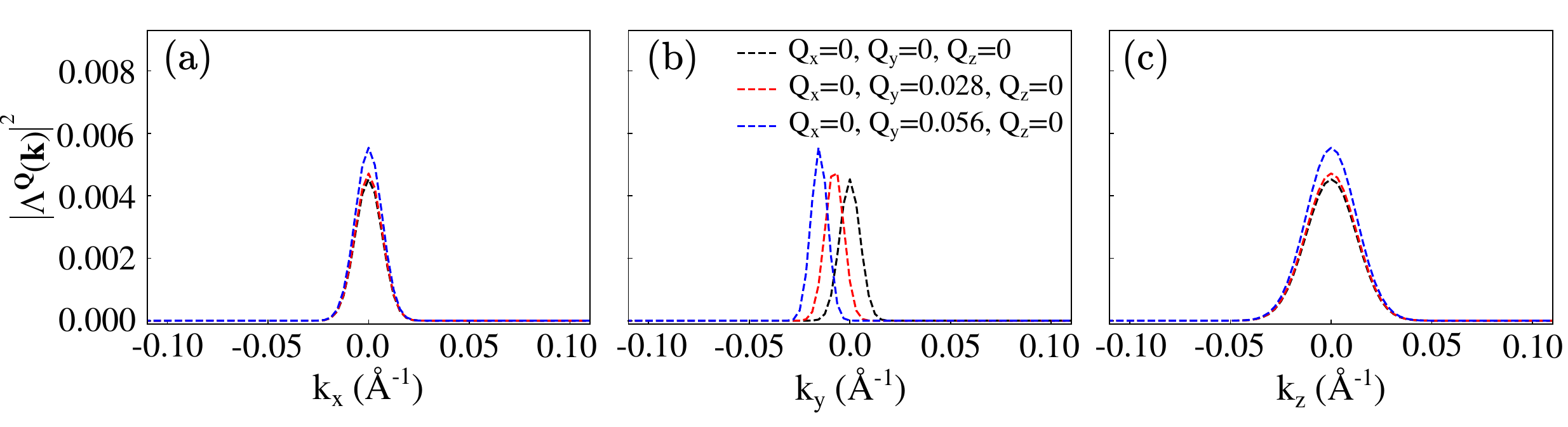}
    \caption{ Exiton's density for different values of $\mathbf{Q}$, 
    (a) along $k_x$ and fixing $k_y=k_z=0$,
    (b) along $k_y$ and fixing $k_x=k_z=0$,
    (c) along $k_z$ and fixing $k_x=k_y=0$ 
    }
    \label{fig:E-WF}
\end{figure}

In Fig.~\ref{fig:E-WF}, we show the exciton wave functions by fixing $Q_x$ and $Q_z$ to zero and displaying different values of $Q_y$.
The wave function is symmetric under changes of sign in the momentum, i.e., $\Lambda^{\mathbf{Q}} \left( \mathbf{k} \right) = \Lambda^{-\mathbf{Q}} \left(\mathbf{k} \right)$.
At this step, we obtained the exciton wave function for a discrete grid of points $\mathbf{Q}$, not far from $\bar{\Gamma}$, and fitted it using a function with the shape
\begin{equation}
    \left| \Lambda^{\mathbf{Q}} (\mathbf{k}) \right|^2 
    = C \prod_{i \in \{x,y,z\}}
    (1 + a_iQ_i) 
    e^{-(Q_i - Q^0_i)^2/\sigma_i^2} \left| \Lambda^{\mathbf{Q}=0} (\mathbf{k}) \right|^2,
\end{equation}
where $C$ and $a_i$ are the parameters that modulate the amplitude, $Q^0_i$ is the position of the maximum of the wave function, and $\sigma_i^2$ is the width of the Gaussian distribution.

\section{Quantum Boltzmann and Exciton Bloch equations}

To characterize the scattering dynamics and momentum redistribution of photo-excited carriers, we model the temporal evolution of exciton populations using a density matrix approach. We define a reduced basis consisting of three primary states: the initial many-body ground state (represented by the occupied valence band at the $\bar{\Gamma}$ point) $| \Psi_0 \rangle$, the optically generated coherent exciton population at the center of mass momentum $\mathbf{Q}=0$ (the "bright" exciton state $\left| \Psi^{\text{exc}}_{\mathbf{Q}=0} \right\rangle$), and the manifold of incoherent exciton populations at $\mathbf{Q} \neq 0$ ($| \Psi^{\text{exc}}_{\mathbf{Q}\neq0} \rangle$). The dynamics are governed by the exciton Bloch equations~\cite{Stefanucci_ExcitonicBlochEquations_2025}, here simplified for the low-density limit. They follow  the same spirit as the semiconductor Bloch equations, where the scattering is described by Markovian Boltzmann-like terms for the incoherent populations.\\

The evolution of the exciton density matrix, $\rho^{\text{exc}}(t)$, is described by the master equation
\begin{equation}
    \frac{d}{dt}\rho^{\text{exc}}(t) = -i \left[ \hat{H}^{\text{exc}}, \rho^{\text{exc}}(t) \right] + S^{\text{exc}}(t),
    \label{eq:density-evolution}
\end{equation}
where the ground state $| \Psi_0 \rangle$ and the bright exciton state $| \Psi^{\text{exc}}_{\mathbf{Q}=0} \rangle$ are coupled by the Hamiltonian term
\begin{equation}
    \left\langle \Psi_0 \left| \hat{H}_{\text{exc}}(t) \right| \Psi^{\text{exc}}_{\mathbf{Q}=0} \right\rangle = - \mathbf{E}(t) \cdot \mathbf{M}_x^{\text{dip}}.
\end{equation}
The pump excitation is provided by a Gaussian electric pulse $\mathbf{E}(t) = E_0 \pmb{\epsilon} e^{-(t-t_0)^2/2\sigma^2} \cos[\omega_{pump}(t-t_0)]$. We employ a pulse with a full-width at half-maximum (FWHM) of $226.27$ fs, a resonant frequency that matches the exciton resonance energy and a field amplitude of $E^{(0)} = 1.5$ kV/cm, which will be explained in detail in the next section. The polarization vector is set to $\pmb{\epsilon} = (0, 0.2912, 0)$, which has been computed using the Fresnel transmission coefficients. The transition matrix elements, $\mathbf{M}_x$, are calculated in the velocity gauge by evaluating the velocity matrix between the valence and conduction states weighted by the exciton wave function
\begin{equation}
    \mathbf{M}_x^{vel} = \frac{1}{\sqrt{N_k}}\sum_{\mathbf{k}} \Lambda^{\mathbf{Q}=0}(\mathbf{k}) \mathbf{v}_{CV} (\mathbf{k}),
\end{equation}
where $\mathbf{v}_{CV} (\mathbf{k}) = -i \left[ \varepsilon_{C}(\mathbf{k}) - \varepsilon_{V}(\mathbf{k})  \right]\mathbf{A}_{CV}(\mathbf{k})$, here $\mathbf{A}_{CV}(\mathbf{k})$ is the Berry connection and then convert it from the velocity to the dipole gauge. 

Coherent excitons at $\mathbf{Q}=0$ scatter to finite momenta $\mathbf{Q} \neq 0$ via interactions with the phonon bath. This process facilitates energy exchange between the electronic system and the lattice.
The momentum-dependent scattering rates, $\Gamma_{\mathbf{Q} \rightarrow \mathbf{Q}'}$, are given by:
\begin{align}
    \Gamma_{\mathbf{Q} \rightarrow \mathbf{Q}'}(T) = 
    2 \pi \gamma^2 
    \left| 
        \mathbf{Q} - \mathbf{Q}'
    \right| 
    & \biggl[ n_{\text{ph}}
    \left( 
        \mathbf{Q} - \mathbf{Q}',T
    \right)
    \delta \Bigl(E_{x} ( \mathbf{Q} ) - E_{x} ( \mathbf{Q}' ) + \omega_{\text{ph}} ( \mathbf{Q}' - \mathbf{Q} ) \Bigr) \nonumber \\ 
    & +
    \left( 
        n_{\text{ph}} (\mathbf{Q} - \mathbf{Q}',T) + 1
    \right) 
    \delta \Bigl(E_{x} ( \mathbf{Q} ) - E_{x} ( \mathbf{Q}' ) - \omega_{\text{ph}} ( \mathbf{Q}' - \mathbf{Q} ) \Bigr)\biggr].
    \label{eq:electron-phonon}
\end{align}
Here, $\omega_{\text{ph}}$ represents the phonon dispersion. In layered compounds, acoustic modes typically exhibit a parabolic dispersion at small wavenumbers; however, finite shear interactions between layers can result in a non-zero slope at $\mathbf{Q}=0$. Consequently, we approximate the acoustic dispersion near the $\bar{\Gamma}$ point linearly as
\begin{equation}
    \omega_{\text{ph}}^2 (\mathbf{Q}) = \sum_{i \in \{x,y,z\}} v_{i}^2 Q_i^2,
\end{equation}
where $v_{i} = \partial_{Q_i} \omega_{\text{ph}}(\mathbf{Q})$ is the group velocity \cite{Hartmut2001,Politano2012}. We adopt group velocities from previous experimental reports \cite{Fujii1982,Guangzhao2015, Zhu2014}. Our model considers only acoustic phonon dispersions, as optical phonons in this system possess energies significantly exceeding the exciton dispersion scale and do not contribute to the scattering channels considered here.

The phonon population, $n_{\text{ph}}(\mathbf{Q},T)$, follows the Bose-Einstein statistics. 
% Variations in temperature modulate the scattering rates and, consequently, the dephasing of the coherent polarization. 
In Eq.~\eqref{eq:density-evolution}, the relaxation and dephasing take the form of  $S^{\text{exc}} = - 2 \sum_{\mathbf{Q}\neq 0} \Gamma_{\mathbf{0}\rightarrow\mathbf{Q}}$.

To ensure the conservation in the density of particles of the momentum grid, we regularize the Dirac delta functions in Eq.~\eqref{eq:electron-phonon} using a Gaussian approximation: $\delta_{\sigma}(\omega) = \frac{1}{\sqrt{2\pi} \sigma}e^{-\omega^2/(2\sigma^2)}$. To avoid the arbitrary tuning of results, we implement a state-dependent broadening width
\begin{equation}
    \sigma_{\mathbf{Q}} = a \left| \frac{\partial E(\mathbf{Q})}{\partial\mathbf{Q}} \right| \Delta Q.
\end{equation}
This approach ensures that the broadening remains proportional to the local energy gradient and grid spacing, $\Delta Q$, thereby guaranteeing a consistent convergence toward the ideal Dirac delta limit as $\Delta Q \rightarrow 0$ \cite{Yates2007,Li2014}.

\section{Connecting theory and experiments}

To connect the theoretical formalism with the experimental results, we follow a systematic procedure to determine the model parameters. This section outlines the formalism for the signal components, the simulation integration, and the extraction of the parameters used: $\alpha$ (LAPE scaling), $E^{(0)}$ (theory electric field amplitude), and $\gamma$ (exciton-phonon coupling strength).

\subsection{Formalism and Simulation Details}
The total trARPES intensity is modeled as the sum of the excitonic signal and the Laser-Assisted Photoemission Effect (LAPE) contribution. The LAPE signal represents a coherent phenomenon where the pump pulse dresses electronic states, appearing as replicas of the valence band shifted by the pump photon energy \cite{LAPE_2006, mahmood_selective_2016, Schueler2022}. It is defined as
\begin{equation}
    I_\text{LAPE}(\mathbf{k}_\parallel,\omega,\tau) = \alpha f^2(\tau)\int dk_z F(k_z - p_\perp) |M_v(\mathbf{k},p_\perp)|^2 \delta(\varepsilon_v(\mathbf{k}) + \omega_{pump} - \omega),
\end{equation}
where $f(\tau)$ is the envelope of the electric field (making $f^2$ the intensity profile), $\omega_{pump}$ is the pump frequency and $\alpha$ is a scaling parameter. 

\subsection{Parameter Determination Procedure}
The parameters are extracted from the experimental data by following the steps:
\begin{itemize}
    \item \textbf{Field Strength ($E^{(0)}$):} We first examine the steady state at late times ($t \to \infty$) after the pump pulse has finished. In this regime, the LAPE signal vanishes. To compare with the experimental distributions, we compute the energy distribution curves (EDC) by integrating the simulated intensity over a momentum range of $[-0.05,0.05]$ \AA$^{-1}$ near $\bar{\Gamma}$. We set the field strength to $E^{(0)} = 1.5$ kV/cm to match the intensity of the exciton peak between the simulation and experiment in the EDC. This value serves as an effective internal field strength.% to reproduce the experimental exciton and should no be taken as the field intensity from the source.
    
    \item \textbf{LAPE Amplitude ($\alpha$) and ROI Analysis:} The parameter $\alpha$ is adjusted by analyzing the "red box" ROI (way below the exciton's binding energy). This region is dominated by LAPE, which is symmetric in time around $\tau=0$. By matching this ROI, we scale it to be $\alpha=7.0 \times 10^{-4}$. The small remaining asymmetry and signal at later times in this region are then used to confirm the minor contribution of the exciton signal in this domain.

    \item \textbf{Effective Coupling Strength ($\gamma$):} Finally, the exciton-phonon coupling strength $\gamma$ is determined by fitting the dynamics in the "green box" and "blue box" ROI . Unlike the LAPE signal, the excitonic signal is asymmetric and exhibits a delayed rise time as bright excitons at $Q=0$ scatter into dark states at $Q \neq 0$. By fitting this rise time and the relative amplitudes across the ROI [Fig. 2b-d], we obtain a unique choice for $\gamma \simeq 2.5\times 10^{-4}$~a.u.
\end{itemize}

\section{Laser Assisted Photoemission Effect}
\begin{figure}[h]
    \centering\includegraphics[width=\linewidth]{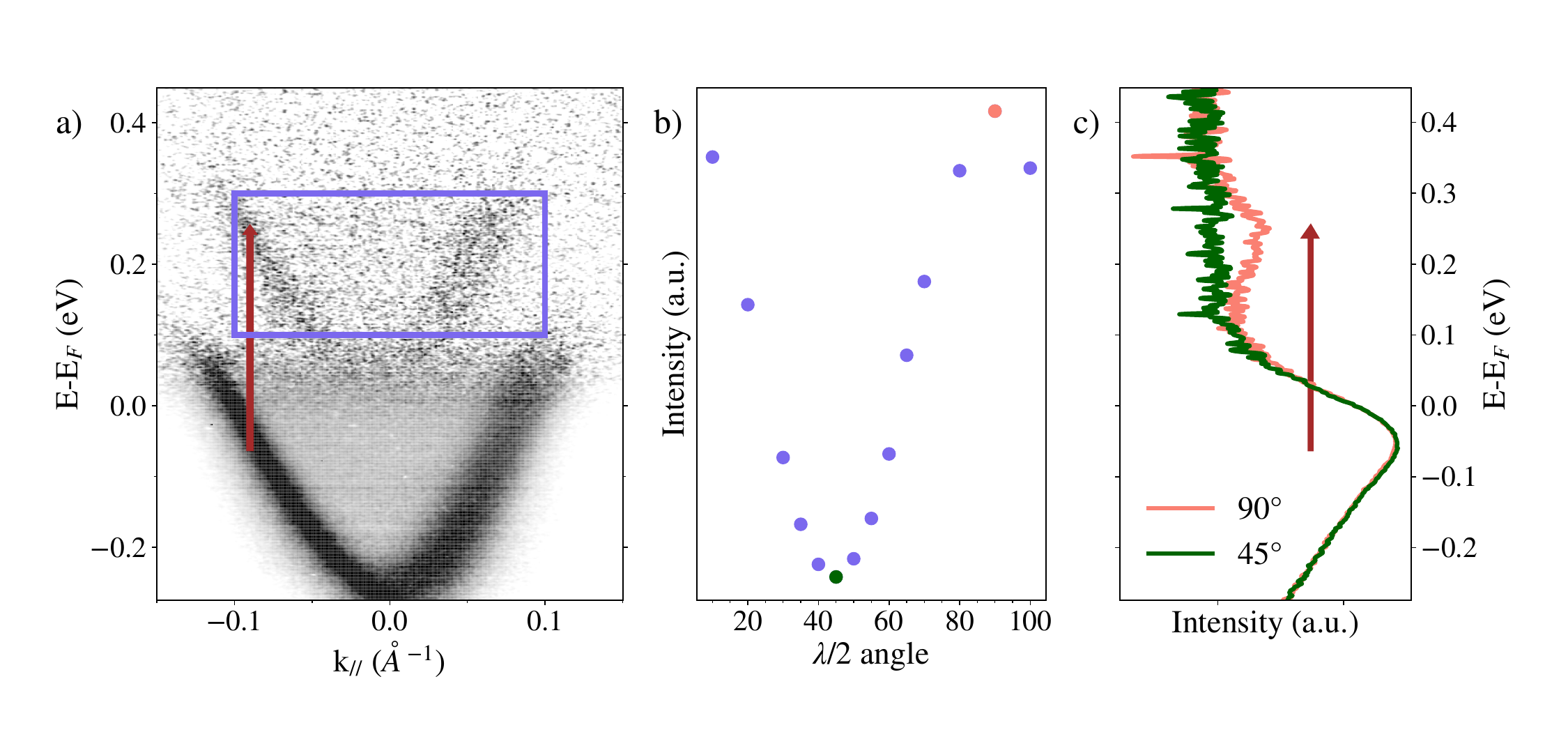}
    \caption{ \textbf{Experimental Measurement of LAPE on Bi$_2$Se$_3$}:
    (a) Pump-probe ARPES spectrum acquired using a $p$-polarized pump with $\hbar\omega_\mathrm{p}=0.31$~eV, showing the replica of the Dirac cone shifted by the photon energy. (b) Integrated signal in the blue ROI of (a) as a function of wave plate angle.  (c) EDCs on pump-probe ARPES spectra using s-polarized (green) and p-polarized (orange) pump photons.
    }
    \label{fig:LAPE_exp}
\end{figure}
In pump-probe ARPES, the laser assisted photoemission effect (LAPE) emerges as a dressing of the final photoelectron state with infrared radiation when using p-polarized pump photons \cite{saathoff_laser-assisted_2008,mahmood_selective_2016}. It manifests as replica of the occupied states, shifted by the pump photon energy $\hbar\omega_\mathrm{p}$, shown in Fig.~\ref{fig:LAPE_exp}(a). To avoid interference with exciton signature measurements, the pump photons are set to s-polarization using a $\lambda/2$ wave plate. The $\lambda/2$ wave plate was calibrated at pump photon energies of $\hbar\omega_\mathrm{p} = 0.31$~eV and $\hbar\omega_\mathrm{p} = 0.41$~eV by performing pump–probe ARPES measurements on the topological insulator Bi$_2$Se$_3$. By selecting a ROI where the replica is expected and integrating its signal across different wave plate configurations, the LAPE intensity is quantified as a function of the $\lambda/2$ angle. The orange EDC in Fig.~\ref{fig:LAPE_exp}(c) shows the maximal LAPE configuration, characterized by a replica band shifted by $0.31$~eV above the main peak. The absence of residual signal in the unoccupied states for the green EDC indicates that the corresponding wave plate setting yields pure s-polarization. The same procedure was performed for 0.41 eV pump photons with the appropriate $\lambda/2$ wave plate.

\end{document}